\documentclass[sn-mathphys-ay]{sn-jnl}

\usepackage{graphicx}%
\usepackage{multirow}%
\usepackage{amsmath,amssymb,amsfonts}%
\usepackage{amsthm}%
\usepackage{mathrsfs}%
\usepackage[title]{appendix}%
\usepackage{xcolor}%
\usepackage{textcomp}%
\usepackage{manyfoot}%
\usepackage{booktabs}%
\usepackage{algorithm}%
\usepackage{algorithmicx}%
\usepackage{algpseudocode}%
\usepackage{listings}%
\usepackage{mathtools}

\theoremstyle{thmstyleone}%
\newtheorem{theorem}{Theorem}
\newtheorem{lemma}[theorem]{Lemma}
\theoremstyle{thmstyletwo}%
\newtheorem{remark}{Remark}%
\newcommand\myeq{\stackrel{\mathclap{\normalfont\mbox{d}}}{=}}
\theoremstyle{thmstylethree}%
\newtheorem{definition}{Definition}%

\begin{document}

\title[Article Title]{Estimation of Spectral Risk Measure for Left Truncated and Right Censored Data}

\author*[]{\fnm{Suparna} \sur{Biswas}\thanks{}}\email{suparnabiswas\_vs@isibang.ac.in}

\author[]{\fnm{Rituparna} \sur{Sen}}\email{rsen@isibang.ac.in}

\affil[]{\orgdiv{Applied Statistics Unit}, \orgname{Indian Statistical Institute}, \orgaddress{\street{Mysore Road}, \city{Bangalore}, \postcode{560059}, \state{Karnataka}, \country{India}}}

\abstract{Left truncated and right censored data are encountered frequently in insurance loss data due to deductibles and policy limits. Risk estimation is an important task in insurance as it is a necessary step for determining premiums under various policy terms. Spectral risk measures are inherently coherent and have the benefit of connecting the risk measure to the user's risk aversion. In this paper we study the estimation of spectral risk measure based on left truncated and right censored data. We propose a non parametric estimator of spectral risk measure using the product limit estimator and establish the asymptotic normality for our proposed estimator. We also develop an Edgeworth expansion of our proposed estimator. The bootstrap is employed to approximate the distribution of our proposed estimator and shown to be second order accurate. Monte Carlo studies are conducted to compare the proposed spectral risk measure estimator with the existing parametric and nonparametric estimators for left truncated and right censored data. Our observation reveal that the proposed estimator outperforms all the estimators for small values of $k$ (coefficient of absolute risk aversion) and for small sample sizes for i.i.d. case. In the dependent case, it demonstrates superior performance for small $k$ across all sample sizes. Finally, we estimate the exponential spectral risk measure for two data sets viz; the Norwegian fire claims and the French marine losses.}


\maketitle









\pagebreak
\section{Introduction}\label{sec1}

Insurance is a data-driven industry that focuses on detecting, quantifying, and protecting against extreme or unexpected events. It is common to encounter incomplete data in the form of truncation and censoring when dealing with insurance data. The most common are the left truncation and right censoring \citep{klugman19}. For example, ordinary insurance deductibles are a common example of left truncation. With typical insurance deductibles, policyholders are not compelled to notify the insurance company of losses that are less than the deductible sum. In this case, the insurer's data is considered truncated because these losses would be covered by the policyholder and not reflected in the insurer's data systems. Right censoring is a common occurrence when working with policy limits. When the actual damages exceed the policy limitations, the insurer's reimbursement shall not exceed the policy limit, and the loss will be deemed right censored. Left truncated data also occurs in banking institutions where similar to insurance deductibles, operational losses below any threshold are not disclosed. The data truncation observed in operational risk was investigated by \cite{Erg16}. The left truncation and right censoring can be defined as follows:\\

\noindent\textbf{Left Truncation:} When left truncation occurs, an observation that is less than or equal to $d$ is not recorded; however, if the observation is more than $d$, it is recorded at the observed value. Let $X$ and $M$ be two random variables, where one represents the size of loss and the other represents the recorded value. Mathematically, it can be represented as:
\begin{equation*} M=\left\{\begin{array}{ll} \text{not\ recorded}& \quad X\leq d \\
X & \quad X>d.\end{array}\right.\end{equation*}

\noindent\textbf{Right Censoring:} In case of right censoring $X$ is observed if it is less than $u$ and if $X$ is higher or equal to $u$ we observe $u$. Mathematically, it can be represented as:
\begin{equation*} M=\left\{\begin{array}{ll} X& \quad X\leq u \\
u & \quad X\geq u.\end{array}\right.\end{equation*}

Measures of financial risk manifest themselves explicitly in many different types of insurance problems, including the setting of premiums and thresholds (e.g.,for deductibles and reinsurance cedance levels) \citep{dowd06}. These measures are used to assess the riskiness of the probability distribution tail, which is an important task in insurance data analytics. The connection between actuarial and financial mathematics is striking here, as premium principles in an actuarial context correspond to risk measures in financial mathematics \citep{pichler2015}. A premium principle is a rule for assigning premiums to the insurance risks. There are various premium principles. For comprehensive discussion see \cite{pichler2015} and \cite{wang1996}. The net-premium principle is often regarded as the most authentic and equitable premium principle in actuarial practice \citep{pichler2015}. An insurance firm using the net-premium concept will inevitably go bankrupt in the long term, even if it covers all its costs by collecting fees from clients. It is essential for the insurance sector to adhere to premium principles to ensure the company's continued existence and ability to provide insurance coverage for clients' future claims. Despite this interesting fact, the exact loss distribution is usually not known. Therefore, alternate premium principles have been created. Also the implementation of an appropriate premium principle or risk measure is a vital concern for the calculation of premium in the insurance industry.\

A simple principle, ensuring risk-adjusted credibility premiums, is the distorted premium principle \citep{wang1996}. This principle is advantageous for insurance firms because actuaries do not need to modify their tools to calculate premiums or reserves. Risk adjusted insurance prices by employing distorted probability measures have been discussed by \cite{wang98}, \cite{heras12} and \cite{Ojeda2023}. The distorted probability relates directly to a special class of risk measures, the spectral risk measures (SRMs) introduced in \cite{acerbi02}. An important study of SRMs, although under the different name distortion functional, was provided in \cite{Tsu09b}. The notions of premium principles through distorting probability measures, distortion functionals, and SRMs are fundamentally similar but differ only in sign conventions, leading to either a concave or convex representation \citep{gzyland08}. The most important premium functional, is the conditional tail expectation (CTE) (in financial context alternative term is expected shortfall (ES)) \cite{heras12}.

Among the different types of risk measures, the coherent risk measures have been traditionally used in actuarial science \citep{Ojeda2023}. \cite{Artz99} established the idea of coherent risk measure (see Appendix). The main characteristic that sets any coherent risk measure apart from other risk measure is subadditivity. We use the SRMs because they are coherent by nature and have the advantage of relating the risk measure to the user's risk aversion \citep{acerbi02}. \cite{acerbi02} introduced the SRM, which is a weighted average of the quantiles of a loss distribution, the weights of which are determined by the user's risk aversion. In addition to that, SRMs are comonotonic and law invariant \citep{gzyland08}. Law invariance is crucial for applications since it is essential for a risk measure to be estimable from empirical data \citep{gzyland08}.

\begin{definition}  Let $\phi\in\mathcal{L}^1([0,1])$ be an admissible risk spectrum (see Appendix) and $X$ be a random loss variable. Then the SRM, according to \cite{gzyland08}, is defined by \begin{equation}\label{S}M_\phi=\int_0^1\phi(u)F^{-1}_X(u)du,\end{equation}
where $\phi$ is called the Risk Aversion Function and $F^{-1}_X$ is the quantile function of $X$(see Appendix).  \end{definition}

According to \cite{gzyland08}, any rational investor can use a new weight function profile $\phi$ to reflect their subjective risk aversion. It is apparent that if $\phi(u)=\frac{1}{1-p}1_{p\leq u\leq 1}$ then $M_\phi$ is the ES which is an SRM. However, value at risk (VaR) is not an SRMM because it is not a coherent risk measure.

The estimation of SRMs is something that interests us. The estimating literature for distortion risk measures (DRMs) is more extensive than that for SRMs. This study focuses on SRM estimation when the data are left truncated and right censored (LTRC). The truncation reduces the sample size of the data and the actual sample size is unknown. Hence, ignoring the truncation would result in inaccurate estimation of SRM which will affect in the assignment of premiums.\

The initial and crucial step in addressing these issues is to obtain point estimates of SRMs and analyze their variability. The most straightforward approach available is the empirical nonparametric technique. But this empirical approach is inefficient due to the scarcity of sample data in the tails \citep{brazauskas2016}. \cite{kaiser2006} discussed that parametric estimators, can considerably increase the efficiency of quantile based estimators. Various authors have studied the estimation methods of DRMs like VaR, ES, proportional hazards transform, Wang transform, and Gini shortfall for LTRC data using the parametric approach. \cite{upretee2020} discussed the empirical nonparametric estimator and parametric estimators of various DRMs for LTRC data and concluded that parametric estimators are more efficient. Again, it is well known that folded distributions are useful models in statistics. \cite{nadarajah2015} introduced several new folded distributions and have discussed an application of the distributions to the left truncated Norwegian fire claims data set. In order to estimate the severity of Norwegian fire claims for the years $1981$ through $1992$, \cite{brazauskas2016} used a number of parametric families.\

But the most significant disadvantage of parametric estimators is that they are susceptible to initial modeling assumptions, resulting in model uncertainty. Therefore, we opt for the nonparametric approach to estimate the SRM, as it is resistant to model risk. Based on the literature, it is evident that the product limit (PL) estimator \citep{tsai1987} is a more suitable approach for handling LTRC data compared to the empirical distribution function. Various authors have studied the estimation of quantile-based risk measures using the PL estimator in the context of LTRC data. \cite{zhou2000} investigated the estimation of a quantile function based on LTRC data by using the kernel smoothing method. \cite{liang2015} studied the conditional quantile estimation based on LTRC data. The authors defined a generalized PL estimator to estimate the conditional quantile. \cite{Shi2018} proposed an improved PL estimator to estimate the quantile function for length biased and right censored data. Again from \cite{cheng2016} we observe that the quantile regression model can also be applied to deal with the LTRC data. But we do not find any literature regarding the estimation of SRM using the PL estimator. When working with LTRC data, it is well understood that the PL quantile estimator is a natural estimator of quantile. Therefore, it is evident that the PL estimator is the most suitable method for estimating SRM in LTRC data \citep{Tsu09b}.\

This paper's objective is to consider the PL estimator in the estimation of $M_{\phi}$ and establish the asymptotic normality of the estimator. Additionally, we approximate the distribution of our proposed estimator using the bootstrap method and derive the Edgeworth expansion. We showed that the bootstrap yields second-order accurate approximations (i.e., with error of size $o(n^{-1/2})$) to the distribution of our proposed estimator. We chose the exponential risk aversion function in our analysis, where the coefficient of absolute risk aversion $k$ plays an important role. Using Monte Carlo (MC) simulations, we compare the accuracy of our proposed estimators for different sample sizes $n$ and $k$. Our simulation study shows that, the proposed estimator outperforms other estimators mentioned in our study for small values of $n$ and $k$ when observations are i.i.d. If observations are dependent it outperforms for all $n$ and $k=5,\ 10$. When $n\geq500$ our proposed estimator outperforms for all values of $k$, except for $k=100$. The paper is arranged as follows. In section 2 we propose the nonparametric estimator of $M_{\phi}$ using the PL estimator. In section 3 we establish the asymptotic normality and derive the Edgeworth expansion of the estimator. Also we investigate Efron's bootstrap method \citep{efron79} to approximate the distribution of the estimator and study its accuracy. In section 4 we compare the finite sample performance of our proposed estimator with various parametric and nonparametric estimator available in the literature for LTRC data using MC simulations and report the findings. Repeated comparisons are made for various sample sizes, estimation methods and for two severity models and dependent case. We also conduct a simulation study, where we estimate the coverage probabilities of confidence intervals. Section 5 offers an empirical analysis based on two data sets:-the Norwegian fire claims and the French marine losses to support the superiority of our approach relative to the existing approaches and compares the riskiness of these two data sets. Finally, in section 6, we discuss the findings and give the concluding remarks.

\section{Proposed estimator}

In this section we briefly discuss the PL estimator under left truncation and right censoring and based on that we define our proposed estimator of SRM. Let ($X,\ T,\ S$) be a random vector where $X$ be the variable of interest with cumulative distribution function (cdf) $F$; $T$ is a random left truncation with cdf $G$ and $S$ is a random right censoring with cdf $L$. We assume that $F$, $G$ and $L$ are continuous. $X$, $T$, $S$ are assumed to be mutually independent. Suppose, one can only observe the pair $Y=X\wedge S$ and $\delta=I(X\leq S)$. Under this restriction, the random variable $X$ is called right censored. Additionally, the triple ($Y, T, \delta$) is called an LTRC observation of $X$ if it is observed only when $Y\geq T$. Nothing is observed if $Y<T$. The conditional distribution
\begin{equation*}P[Y\leq y,T\leq t,\delta=q|T\leq Y],\text{for}\ y,\ t\in(-\infty,\infty),\ q=0\ \text{or}\ 1,  \end{equation*}
defines the LTRC model. Let $\alpha=P(Y\geq T)>0$, and let $W$ denote the cdf of $Y$. If $X$ is independent of $S$ then $1-W=(1-F)(1-L)$.

Let ($X_i,\ T_i,\ S_i,\ \delta_i$), $1\leq i\leq N$ be a sequence of random vectors which are i.i.d. The size of the sample $n$ that was actually observed is random as a result of truncation, where $n\leq N$ and $N$ unknown. But we can regard the observed sample, ($X_i,\ T_i,\ S_i,\ \delta_i$), $1\leq i\leq n$ as being generated by independent random variables $X_i$, $T_i$, $S_i$, $1\leq i\leq N$. From the SLLN, $\frac{n}{N}\rightarrow\alpha=P(Y\geq T)>0\ a.s.$

Now given the value of $n$, the data ($X_i, T_i, S_i, \delta_i$) are still i.i.d. That is, ($Y_i, T_i, \delta_i$), $i=1,2,\cdots,n$ are i.i.d random samples of ($Y,T,\delta$) which is observed, but the joint distribution of $Y$ and $T$ becomes
$$H^*(y,t)=P(Y\leq y,T\leq t|T\leq Y).$$
\cite{wood1985} pointed out that $F$ is identifiable only if certain requirements on the support of $F$ and $G$ are met. For any df $K$, let \begin{equation*} a_K=\inf\{t:K(t)>0\} \end{equation*} and \begin{equation*} b_K=\sup\{t:K(t)<1\} \end{equation*} denote the left- and right-endpoints of its support. For the LTRC data, $F$ is identifiable if $a_G\leq a_W$ and $b_G\leq b_W$.

Let \begin{align}\label{C}C(z)=P(T\leq z\leq Y|T\leq Y)=\alpha^{-1}G(z)(1-W(z-))=\alpha^{-1}P(T\leq z\leq S)(1-F(z-)),\end{align}
where $a_W\leq z<\infty$. $C$ can be estimated by its empirical estimator
$$C_n(z)=n^{-1}\sum_{i=1}^nI(T_i\leq z\leq Y_i).$$

Let $b$ denote a constant less than $b_W$. From \cite{stute1993} we observe that though $C$ is strictly positive on $a_G<z<b_W$, $C_n$ may vanish for some $z\in(a_G,\ b_W)$. In particular, $C_n$ vanishes for all $z$ less than the smallest $T$ order statistics. 

It is critical for LTRC data to be able to generate nonparametric estimates of various features of the cdf $F$. \cite{zhou2000} defined the nonparametric maximum likelihood estimator of $F$, called the PL estimator, as

\begin{equation} \label{F12} \widehat{F}(x)=\left\{ \begin{array}{ll}1-\prod_{Y_{i}\leq x}\Big[\frac{nC_n(Y_{i})-1}{nC_n(Y_{i})}\Big]^{\delta_{i}}& \mathrm{for}\quad x<Y_{(n)} \\
1 & \mathrm{for}\quad x\geq Y_{(n)},\end{array}\right.\end{equation}

\noindent where $Y_{(n)}=\max(Y_1,\ldots, Y_n)$. The properties of $\widehat{F}$ have been studied by \cite{wang1987}, \cite{lai1991}, \cite{stute1993}, \cite{zhou1996}, and \cite{zhou1999}. Again, \cite{uzunog1992}, \cite{gijbels1993}, \cite{gu1995}, \cite{sun1997}, and \cite{sun1998} investigated nonparametric estimates of $F$'s density and hazard rate. Based on equation (\ref{F12}) we propose the following estimator for $M_{\phi}$,

\begin{align}\label{Mt}\widehat{M}_{Prod}=\int_0^1\phi(u)\widehat{F}^{-1}(u)du. \end{align}

\noindent The asymptotic normality and the Edgeworth expansion of $\widehat{M}_{Prod}$ is established in the following section.

\section{Distribution of the estimator}

In order to find the confidence intervals of the point estimates of SRM. We need to know the distribution of our proposed SRM estimator. In this section we prove that our SRM estimator asymptotically follows normal distribution and further improve the normal approximations to higher-order by deriving the Edgeworth expansion of our estimator $\widehat{M}_{Prod}$. Again, the bootstrap is employed to approximate the distribution of our proposed estimator $\widehat{M}_{Prod}$. The method employed in our derivation of the desired Edgeworth expansion is first to approximate the $\widehat{M}_{Prod}$ by a $U$-statistic with some sufficiently small error term and then to apply the known results of the Edgeworth expansion for $U$-statistic by checking the relevant conditions \citep{wang06}.

\subsection{Asymptotic normality}
In the current model, as discussed by \cite{zhou1996}, let us consider the following assumptions:\

\noindent\emph{Assumption 1:} $a_G\leq a_W$, $b_G\leq b_W$.

\noindent\emph{Assumption 2:} $\int_{a_W}^\infty\frac{dF(z)}{G^2(z)}<\infty$.
Let \begin{align*} W^*(y)=P(Y\leq y,\delta=1|T\leq Y),\end{align*} and
\begin{align*} W_{n}^*(y)=n^{-1}\sum_{i=1}^{n}I(Y_i\leq y,\delta_i=1). \end{align*}

\begin{theorem}\label{theo2}
Let assumptions 1, 2, and $a_W\leq F^{-1}(t)\leq b<b_W$ are satisfied. Then
$$\sqrt{n}(\widehat{M}_{Prod}-M_\phi)\xrightarrow{\text{d}}N(0,\sigma^2)$$
where \begin{equation}\label{sig}\sigma^2=\int_0^1\int_0^1\frac{(1-u)(1-v)}{f(F^{-1}(u))f(F^{-1}(v))}\Big[\int_{a_W}^{F^{-1}(u)}\frac{dW^*(y)}{C^2(y)}\int_{a_W}^{F^{-1}(v)}\frac{dW^*(y)}{C^2(y)}-uv\Big]\phi(u)\phi(v)dudv. \end{equation}
\end{theorem}

\noindent\textbf{Proof:} See Appendix.\

\noindent Hence, from Theorem \ref{theo2} we can say that our proposed estimator $\widehat{M}_{Prod}$ is asymptotically distributed with Gaussian behavior as the sample size goes up.

\begin{remark} \cite{gijbels1993} density estimate can be used to replace the density $f$, and empirical estimates of all other unknown values can be used to replace their estimates to provide a consistent estimate of $\sigma^2$ \citep{zhou2000}. \end{remark}

\subsection{Edgeworth Expansion}

For simplicity, we shall denote
\begin{equation*}
V_i=\{Y_i,\ T_i,\ \delta_i\},\ i=1,\cdots,n.
\end{equation*}
For $i,\ j=1,\cdots,n$, let

\begin{eqnarray*}
\zeta(V_i\ ;\ x)&=&\frac{I[Y_i\leq x,\delta_i=1]}{C(Y_i)}-\int_{a_W}^x\frac{I[T_i\leq y\leq Y_i]dW^*(y)}{C^2(y)} \\
\psi_1(V_i,\ V_j\ ;\ x)&=&\frac{-I[Y_i\leq x]I[T_j\leq Y_i\leq Y_j]}{C^{2}(Y_i)} \\
\psi_2(V_i,\ V_j\ ;\ x)&=&\int_{a_W}^x\frac{I[T_i\leq y\leq Y_i]I[T_j\leq y\leq Y_j]dW^*(y)}{C^3(y)} \\
\psi(V_i,\ V_j\ ;\ x)&=&\psi_1(V_i,\ V_j\ ;\ x)\delta_i+\psi_2(V_i,\ V_j\ ;\ x)-E[\psi_1(V_i,\ V_j\ ;\ x)+\psi_2(V_i,\ V_j\ ;\ x)|V_i] \\
h_1(V_i,\ V_j\ ;\ x)&=&\zeta(V_i\ ;\ x)+\zeta(V_j\ ;\ x)-\zeta(V_i\ ;\ x)\zeta(V_j\ ;\ x)+\psi(V_i,\ V_j\ ;\ x)+\psi(V_j,\ V_i\ ;\ x) \\
\sigma_0^2&=&\int_{a_W}^x\frac{dW^*(y)}{C^2(y)} \\
\sigma_1^2&=&(1-x)^2\int_{a_W}^{F^{-1}(x)}\frac{dW^*(u)}{C^2(u)} \\
\end{eqnarray*}
and \begin{equation*}
U^{(n)}=\frac{1}{n^2}\sum_{i<j}h_1(V_i,\ V_j\ ;\ x).
\end{equation*}
It is easy to see that $U^{(n)}$ is a $U$-statistic with a symmetric kernel. Theorem \ref{theo6.5} gives the $U$-statistic representation of our proposed estimator.
\begin{theorem}\label{theo6.5}
Let, assumptions 1, 2, and $a_W\leq F^{-1}(t)\leq b< b_W$ are satisfied. Then
\begin{equation*}
\widehat{M}_{Prod}-M_{\phi}=-\int_0^1\frac{1}{f(F^{-1}(u))}\Big[\frac{(1-u)}{n^2}\sum_{i<j}h_1(V_i;\ V_j;\ F^{-1}(u))-\frac{1-u}{2n}\int_{a_W}^{F^{-1}(u)}\frac{dW^*(x)}{C^2(x)}\Big]\phi(u)du+\alpha_n
\end{equation*}
with $P(\sqrt{n}|\alpha_n|>cn^{-1/2}\log{n}^{(-1)})=o(n^{-1/2})$.
\end{theorem}

\noindent\textbf{Proof:} See Appendix.\

\noindent Now, similar to Lemma \ref{lem3.5} in Appendix we obtain another lemma stated below.
\begin{lemma}\label{lem3.7} Under the conditions of Theorem \ref{theo6.5}, for any $a_W<F^{-1}(t)<b_W$, we have
\begin{equation*}\sup_y|P(\sqrt{n}\sigma_{01}^{-1}\frac{1}{n^2}\sum_{i<j}h_1(V_i,V_j;F^{-1}(x))\leq y)-E_{n1}(y)|=o(n^{-1/2}),\end{equation*}
where
\begin{equation*}
E_{n1}(y)=\Phi(y)-\frac{\kappa_3}{6}n^{-1/2}\phi_0(y)(y^2-1),\ \kappa_3=\frac{1}{\sigma_{01}^3}\Big(-\frac{15}{2}\sigma_{01}^4+\int_{a_W}^{F^{-1}(x)}\frac{dW^*(t)}{C^3(t)}\Big),
\end{equation*}
\begin{equation*}
\sigma_{01}^2=\int_{a_W}^{F^{-1}(x)}\frac{dW^*(t)}{C^2(t)},
\end{equation*}
$\Phi$ is the standard normal distribution function, and $\phi_0$ is the standard normal density function.
\end{lemma}

\noindent\textbf{Proof:} The proof is exactly similar to that of Lemma 3 in \cite{chang91} and hence omitted.

\begin{theorem}\label{theo3.8} Let assumptions 1, 2, and $a_W\leq F^{-1}(t)\leq b<b_W$ are satisfied. Then
\begin{equation*}
\sup_y|P(\sqrt{n}\sigma^{-1}(\widehat{M}_{Prod}-M_\phi)\leq y)-E_{n2}(y)|=o(n^{-1/2}),
\end{equation*}
\begin{equation*}
E_{n2}(y)=\Phi(y)-n^{-1/2}\phi_0(y)[\frac{\kappa_3}{6}(y^2-1)+\frac{1}{2}\sigma_{01}]
\end{equation*}
\end{theorem}

\noindent\textbf{Proof:} See Appendix.\

Hence, Theorem \ref{theo3.8} gives the Edgeworth expansion of our proposed estimator $\widehat{M}_{Prod}$.

\subsection{Bootstrap Approximation}
In this section we conside the Efron's bootstrap \citep{efron79} to obtain ``good'' approximations to the distribution of our proposed estimator $\widehat{M}_{Prod}$. To describe the bootstrap with LTRC, let ($Y_i^*$, $T_i^*$, $\delta_i^*$), $i=1,\cdots,n$, be a random sample obtained by drawing with replacement from the observations \{$V_i$\}, $i=1,\cdots,n$. Let $P^*$ denote the bootstrap probability, given the sample \{$V_i$\}, $i=1,\cdots,n$. Let,
\begin{equation*}
C_n^*=\frac{1}{n}\sum_{i=1}^nI[T_i^*\leq z\leq Y_i^*],
\end{equation*}
\begin{equation*}
W_n^{*^*}(y)=\frac{1}{n}\sum_{i=1}^nI[Y_i^*\leq y,\delta_i^*=1].
\end{equation*}

Then, the bootstrap PL estimator $\widehat{F}^*(x)$ can be defined by
\begin{equation*} \widehat{F}^*(x)=\left\{ \begin{array}{ll}1-\prod_{Y_{i}^*\leq x}\Big[\frac{nC_n^*(Y_{i}^*)-1}{nC_n^*(Y_{i}^*)}\Big]^{\delta_i^*}& \mathrm{for}\quad x<Y_{(n)}^* \\
1 & \mathrm{for}\quad x\geq Y_{(n)}^*,\end{array}\right.\end{equation*}
where $Y_{(n)}^*=\max(Y_1^*,\ldots, Y_n^*)$. The bootstrap version of our proposed estimator $\widehat{M}_{Prod}^*$ is given by
\begin{equation*}
\widehat{M}_{Prod}^*=\int_0^1\phi(u)\widehat{F}^{*^{-1}}(u)du,
\end{equation*}
and $\sigma^{*^2}$ is the bootstrap variance.
To investigate the performance of the bootstrap approximation, we shall first establish the Edgeworth expansion for the distribution of the bootstrap estimator $\widehat{M}_{Prod}^*$ in our next theorem.

\begin{theorem}\label{theo3.9}
Under the assumption of Theorem (\ref{theo3.8}), we have with probability 1 for any $a_W\leq F^{-1}(t)\leq b<b_W$
\begin{equation*}
\sup_y|P^*(\sqrt{n}\sigma^{*^{-1}}(\widehat{M}_{Prod}^*-\widehat{M}_{Prod})\leq y)-P(\sqrt{n}\sigma^{-1}(\widehat{M}_{Prod}-M_\phi)\leq y)|=o(n^{-1/2}).
\end{equation*}
\end{theorem}

\noindent\textbf{Proof:} See Appendix.\

\begin{remark}In Theorem \ref{theo3.9} we prove two results. First we prove that the distribution of the bootstrap $\widehat{M}_{Prod}^*$ estimator approximates to the bootstrap version of $E_{n2}^*$ almost surely with an error term $o(n^{-1/2})$ (i.e. $\sup|P^*(\cdot)-E_{n2}^*|=o(n^{-1/2})$). Finally, we show that the bootstrap version of the $E_{n2}^*$ approximates to $E_{n2}$ almost surely with an error term $o(n^{-1/2})$. Altogether Theorem \ref{theo3.9} means that, the distribution of the bootstrap $\widehat{M}_{Prod}^*$ estimator approximates its true distribution with an error term $o(n^{-1/2})$. By this result, the coverage error for the bootstrap confidence interval will be $o(n^{-1/2})$.

\end{remark}

\section{Simulations}

In this section we study the impact of unreported events such as insurance deductibles and policy limits on the estimation of SRM. A Monte Carlo study is carried out to provide some small sample comparisons. We compare our proposed estimator $\widehat{M}_{Prod}$ with the nonparametric and parametric estimators mentioned in \cite{upretee2020}. First we define the two severity distributions mentioned in \cite{upretee2020}.

\subsection{Independent case}\label{Ex}

Let the random sample $X^*_1,\cdots,X^*_n$ satisfy the following conditional event:
\begin{equation}\label{X*}X^*_i\myeq\{\min(X_i,u)|X_i>d\},\ i=1,\cdots,n \end{equation}
where $d$ is the deductible and $u$ is the policy limit. The distribution function is given by
\begin{equation*}F_*(x)=\left\{ \begin{array}{ll}0& \mathrm{for}\quad x\leq d \\
\frac{F(x)-F(d)}{1-F(d)} & \mathrm{for}\quad d<x<u \\
1 & \mathrm{for}\quad x\geq u.\end{array}\right. \end{equation*}

 Two severity distributions used in our simulation study are defined as follows:
\begin{enumerate}
\item \textbf{Shifted Exponential Distribution}: Let, the random variable $X$ follows a shifted exponential distribution with a location (shift) parameter $x_0>0$ and scale parameter $\theta>0$ and denote it as $X\sim \mathrm{Exp}(x_0,\theta)$. The df is defined as
    \begin{align*} F(x)=1-e^{-(x-x_0)/\theta},\ x\geq x_0. \end{align*}
    For $X\sim \mathrm{Exp}(x_0,\theta)$, we have $[F(x)-F(d)]/[1-F(d)]=1-e^{-(x-d)/\theta}$ for $d<x<u$ ($d>x_0$) and the quantile of $X^*$ is given by
    \begin{equation*}F^{-1}_*(p)=\left\{ \begin{array}{ll}-\theta \mathrm{log}(1-p)+d& \mathrm{for}\quad 0\leq p<1-e^{-(u-d)/\theta}, \\
    u & \mathrm{for}\quad 1-e^{-(u-d)/\theta}\leq p\leq1.\end{array}\right.  \end{equation*}

\item \textbf{Pareto I Distribution}: Let, the random variable $X$ follows a Pareto I distribution with a scale parameter $x_0>0$ and shape parameter $\alpha>0$ and denote it as $X\sim \mathrm{PaI}(x_0,\alpha)$. The df is defined as
    \begin{align*}F(x)=1-(x_0/x)^{\alpha},\ x\geq x_0. \end{align*}
    For $X\sim \mathrm{PaI}(x_0,\alpha)$, we have $[F(x)-F(d)]/[1-F(d)]=1-(d/x)^{\alpha}$ for $d<x<u$ ($d>x_0$) and the quantile of $X^*$ is given by
     \begin{equation*}F^{-1}_*(x)=\left\{ \begin{array}{ll}d(1-p)^{-1/\alpha}& \mathrm{for}\quad 0\leq p<1-(d/u)^{\alpha}, \\
    u & \mathrm{for}\quad 1-(d/u)^{\alpha}\leq p\leq1.\end{array}\right.  \end{equation*}
\end{enumerate}

\noindent We consider the truncation and censoring thresholds as mentioned in \cite{upretee2020}:
\begin{itemize}
\item $d=4.10^3$ (corresponds to the $95\%$ data truncation under $\mathrm{Exp}(x_0=10^3,\theta=10^3)$ and $93.8\%$ under $\mathrm{PaI}(x_0=10^3,\alpha=2.0)$).
\item $u=14.10^3$ (corresponds to the $0.0045\%$ data censoring under $F_E=\mathrm{Exp}(d=4.10^3,\theta=10^3)$ and $8.2\%$ under $F_P=\mathrm{PaI}(d=4.10^3,\alpha=2.0)$).
\end{itemize}

Next, we define different nonparametric and parametric estimators of SRM. We use the exponential risk aversion function defined by \cite{cotter06},

\begin{align}\label{phi}\phi(u)=\frac{ke^{-k(1-u)}}{1-e^{-k}}.  \end{align}

where $k\in(0, \infty)$ is the user's coefficient of absolute risk aversion and hence we call it an exponential SRM. The absolute risk aversion coefficient $k$, like the confidence level in the VaR and ES, plays a significant part in SRMs. According to \cite{cotter06},  the higher the value of $k$, the more we pay attention to higher losses in comparison to other losses. Hence our choice of $k$ includes 1, 5, 10, 20, 100, and 200.

\subsubsection{Parametric estimators}
Two parametric estimators are defined here.
\begin{enumerate}
\item \textbf{Maximum Likelihood(ML) estimation}\citep{upretee2020} If $X\sim \mathrm{Exp}(x_0,\theta)$ i.e. Shifted exponential distribution then the estimate of VaR is given by
\begin{align*}\mathrm{VaR}_{p,ML}=x_0-\hat{\theta}_{ML}\mathrm{log}(p), \end{align*}
where $$\hat{\theta}_{ML}=\frac{\sum_{i=1}^n(x^*_i-d)1_{\{d<x^*_i<u\}}+(u-d)1_{\{x^*_i=u\}}}{\sum_{i=1}^n1_{\{d<x^*_i<u\}}}.$$
If $X\sim \mathrm{PaI}(x_0,\alpha)$ i.e. Pareto I distribution then the estimate of VaR is given by
\begin{align*}\mathrm{VaR}_{p,ML}=x_0p^{-1/\hat{\alpha}_{ML}}, \end{align*}
where
$$\hat{\alpha}_{ML}=\frac{\sum_{i=1}^n1_{\{d<x^*_i<u\}}}{\sum_{i=1}^n\mathrm{log}(x^*_i/d)1_{\{d<x^*_i<u\}}+\mathrm{log}(u/d)\sum_{i=1}^n1_{\{x^*_i=u\}}}.$$
Then the exponential SRM estimator is given by
\begin{align}\label{ML1} \widehat{M}_{ML}=\int_0^1\phi(u)\mathrm{VaR}_{u,ML}du,\end{align} where $\phi$ is given in (\ref{phi}).

\item \textbf{Percentile Matching(PM) estimation}\citep{upretee2020} If $X\sim \mathrm{Exp}(x_0,\theta)$ then the estimate of VaR is given by
\begin{align*}\mathrm{VaR}_{p,PM}=x_0-\hat{\theta}_{PM}\mathrm{log}(p), \end{align*}
where $$\hat{\theta}_{PM}=\frac{\theta-x^*_{[np_1]}}{\mathrm{log}(1-p_1)},\ for\ 0<p_1<1-e^{-(u-d)/\theta}.$$

If $X\sim \mathrm{PaI}(x_0,\alpha)$ then the estimate of VaR is given by
\begin{align*}\mathrm{VaR}_{p,PM}=x_0p^{-1/\hat{\alpha}_{PM}}, \end{align*}
where
$$\hat{\alpha}_{PM}=\frac{\mathrm{log}(1-p_1)}{\mathrm{log}(d/x^*_{[np_1]})},\ for\ 0<p_1<1-(d/u)^{\alpha}.$$

Then the exponential SRM estimator is given by
\begin{align}\label{PM1} \widehat{M}_{PM}=\int_0^1\phi(u)\mathrm{VaR}_{u,PM}du, \end{align} where $\phi$ is given in (\ref{phi}).
\end{enumerate}

\subsubsection{Nonparametric estimators}
Two nonparametric estimators are defined here.
\begin{enumerate}
\item \textbf{Empirical(EMP)}\citep{upretee2020} For $0<p<\frac{F(x)-F(d)}{1-F(d)}\leq1$, the empirical estimator of VaR is
    \begin{align*}\mathrm{VaR}_{p,EMP}=x^*_{[np]}.  \end{align*}
    Then the exponential SRM estimator is given by
    \begin{align}\label{EMP} \widehat{M}_{EMP}=\int_0^1\phi(u)\mathrm{VaR}_{u,EMP}du, \end{align} where $\phi$ is given in (\ref{phi}).

\item \textbf{Kernel estimator} The kernel quantile estimator for LTRC data proposed by \cite{zhou2000} is defined as follows
\begin{align*} \hat{Q}(t)=\frac{1}{h}\int_0^1\widehat{F}^{-1}K\left(\frac{x-t}{h}\right)dx, \end{align*} where $h$ is the bandwidth and $K$ is the kernel function. Then the kernel based estimator of exponential SRM is defined as follows
\begin{align}\label{Q1} \widehat{M}_{\hat{Q}}=\int_0^1\phi(u)\hat{Q}(u)du, \end{align} where $\phi$ is given in (\ref{phi}). The bandwidth $h=0.4$ is chosen based on the paper by \cite{liang2015} and the kernel function is chosen as Epanechnikov kernel.
\end{enumerate}

\subsubsection{Simulation setting and results}

From a specified LTRC severity distribution, we generate $10000$ samples of a specified length $n=30,\ 100,\ 500$ and for $k=1,\ 5,\ 10,\ 20,\ 100,\ 200$. For each sample, we estimate the exponential SRM estimators defined in equation (\ref{Mt}), (\ref{ML1}), (\ref{PM1}), (\ref{EMP}) and (\ref{Q1}). Then, based on those $10000$ SRM estimates, we compute their mean, standard deviation (SD) and root mean squared error (RMSE). In Table \ref{tablesrm2} and \ref{tablesrm3} we report the means, SDs and RMSEs of the exponential SRM estimates for the two distributions $F_E$ and $F_P$. All entries are measured in 1000's. In Figure 1 and 2 we plot the logarithm of ratio of the RMSEs (RMSE of all the estimator is divided by the RMSE of the $\widehat{M}_{Prod}$ estimator) of different exponential SRM estimators for $F_E$ and $F_P$ distribution. From Table 1 and Figure 1 we observe that for $n=30$ and for all values of $k$ except for $k=20$ the $\widehat{M}_{Prod}$ estimator outperforms other estimators. For $n=100$ and for $k=1$ and $5$ the $\widehat{M}_{Prod}$ estimator outperforms all the estimators. When $n=500$ and $k=200$ the $\widehat{M}_{Prod}$ estimator outperform other estimators, but for other values of $k$ we see that $\widehat{M}_{ML}$ estimator outperform other estimators. Again, for $k=20$ and for all the sample sizes the $\widehat{M}_{ML}$ estimator also outperforms all the estimators. We also observe that when $k=100,\ 200$ and for all the sample sizes the $\widehat{M}_{EMP}$ estimator outperforms all the estimator. From Table 2 and Figure 2 we observe that for $n=30$ and $k=1,\ 5$ the $\widehat{M}_{Prod}$ estimator outperforms all the estimator in case of $F_P$ distribution. We even observe that for $n=100,\ 500$ and for all values of $k$ except for $k=10$, $\widehat{M}_{Prod}$ estimator outperforms all the estimator. We also observe that when $n=30$ and $k=10,\ 20,\ 100,\ 200$ the $\widehat{M}_{EMP}$ estimator outperforms other estimator. We observe that when $n=500$ the MSEs of $\widehat{M}_{Prod}$ and $\widehat{M}_{PM}$ estimator are very close to each other in case of $F_P$ distribution.\

\begin{table}[htbp]
\centering
\caption{Mean, SD \& RMSE of exponential SRM estimators for $F_E=Exp(x_0=10^3, \theta=10^3)$ distribution.}

\begin{tabular}{|l|l|l|l|l|l|l|l|l|}
\hline
n & Estimation Methods & Estimated Values & $k=1$ & $k=5$ & $k=10$ & $k=20$ & $k=100$ & $k=200$ \\ [0.5ex]
\hline
     & Theoretical Values& & $2.260$ & $3.203$ & $3.878$ & $4.572$ & $6.182$ & $6.876$ \\
\hline
$30$ & $\widehat{M}_{ML}$ & Mean & $2.228$ & $3.115$ & $4.003$ & $4.465$ & $6.368$ & $7.221$ \\
    && SD & $0.294$ & $0.267$ & $0.578$ & $0.726$ & $1.098$ & $1.279$ \\
    &&RMSE & $0.2957$ & $0.2811$& $0.5914$& $0.7338$& $1.1136$ & $1.3247$ \\
    &$\widehat{M}_{PM}$& Mean & $2.177$ & $3.253$ & $3.539$ & $4.199$ & $6.217$ & $6.653$ \\
    && SD & $0.374$ & $0.524$ & $0.519$ & $0.630$ & $1.115$ & $0.986$  \\
    &&RMSE& $0.3831$ & $0.5265$ & $0.6199$ & $0.7321$ & $1.1155$ & $1.0109$ \\
    &$\widehat{M}_{EMP}$& Mean & $4.716$ & $4.720$ & $4.811$ & $5.713$ & $6.012$ & $6.562$ \\
    && SD & $0.188$ & $0.179$ & $0.182$ & $0.183$ & $0.436$ & $0.603$ \\
    && RMSE& $2.4632$ & $1.5285$& $0.9506$ & $1.1556$ & $0.4680$ & $0.6429$ \\
    &$\widehat{M}_{Prod}$& Mean & $2.202$ & $3.310$ & $4.121$ & $4.991$ & $6.070$ & $6.768$ \\
    && SD & $0.252$ & $0.247$ & $0.449$ & $0.648$ & $0.950$ & $1.050$ \\
    &&RMSE& $0.2586$& $0.2692$ & $0.5105$ & $0.7717$ & $0.9566$ & $1.0555$ \\
    &$\widehat{M}_{\hat{Q}}$ & Mean & $5.899$ & $7.966$ & $8.707$ & $9.110$ & $9.396$ & $9.430$\\
    && SD & $0.145$ & $0.078$ & $0.075$ & $0.099$ & $0.111$ & $0.122$ \\
    &&RMSE& $3.6419$ & $4.7636$ & $4.8291$& $4.5391$ & $3.2159$ & $2.5569$ \\
$100$ & $\widehat{M}_{ML}$ & Mean & $2.237$ & $3.177$ & $3.861$ & $4.701$ & $6.190$ & $6.779$ \\
    && SD & $0.201$ & $0.187$ & $0.108$ & $0.299$ & $0.558$ & $0.572$  \\
    &&RMSE & $0.2023$& $0.1888$ & $0.1093$ & $0.3256$ & $0.5581$ & $0.5798$ \\
    &$\widehat{M}_{PM}$& Mean & $2.228$ & $3.148$ & $3.828$ & $4.508$ & $6.053$ & $6.729$ \\
    && SD & $0.135$ & $0.235$ & $0.396$ & $0.492$ & $0.553$ & $0.627$  \\
    &&RMSE& $0.1387$ & $0.2414$ & $0.3991$ & $0.4961$ & $0.5678$ & $0.6421$ \\
    &$\widehat{M}_{EMP}$& Mean & $4.698$ & $4.699$ & $4.700$ & $4.896$ & $6.047$ & $6.726$  \\
    && SD & $0.095$ & $0.099$ & $0.105$ & $0.095$ & $0.263$ & $0.372$  \\
    &&RMSE&$2.4399$ & $1.4993$ & $0.8287$ & $0.3376$ & $0.2956$ & $0.4011$ \\
    &$\widehat{M}_{Prod}$& Mean & $2.312$ & $3.220$ & $4.019$ & $4.020$ & $6.020$ & $6.919$  \\
    && SD & $0.101$ & $0.130$ & $0.304$ & $0.346$ & $0.499$ & $0.580$  \\
    &&RMSE& $0.1136$& $0.1311$ & $0.3351$ & $0.6515$ & $0.5246$ & $0.5816$ \\
    &$\widehat{M}_{\hat{Q}}$ & Mean & $5.789$ & $7.824$ & $8.638$ & $8.992$ & $9.246$ & $9.299$ \\
    && SD & $0.015$ & $0.016$ & $0.033$ & $0.017$ & $0.028$ & $0.018$ \\
    && RMSE & $3.5290$ & $4.6210$ & $4.7601$ & $4.4200$ & $3.0641$ & $2.4231$\\
$500$& $\widehat{M}_{ML}$ & Mean & $2.276$ & $3.154$ & $3.882$ & $4.563$ & $6.316$ & $6.765$  \\
    && SD & $0.042$ & $0.105$ & $0.093$ & $0.124$ & $0.176$ & $0.315$ \\
    &&RMSE& $0.0422$ & $0.1159$ & $0.0931$ & $0.1243$ & $0.2212$ & $0.3340$ \\
    &$\widehat{M}_{PM}$& Mean & $2.273$ & $3.212$ & $3.827$ & $4.588$ & $6.204$ & $6.901$  \\
    && SD & $0.068$ & $0.089$ & $0.175$ & $0.145$ & $0.210$ & $0.238$  \\
    &&RMSE & $0.0692$ & $0.0895$ & $0.1823$ & $0.1459$ & $0.2111$ & $0.2393$ \\
    &$\widehat{M}_{EMP}$& Mean & $4.596$ & $4.609$ & $4.695$ & $5.201$ & $6.078$ & $6.766$  \\
    && SD & $0.044$ & $0.044$ & $0.048$ & $0.043$ & $0.119$ & $0.172$  \\
    &&RMSE& $2.3364$ & $1.4067$ & $0.8184$ & $0.6305$ & $0.1580$ & $0.2042$ \\
    &$\widehat{M}_{Prod}$& Mean & $2.308$ & $3.284$ & $3.909$ & $4.612$ & $6.283$ & $6.819$  \\
    && SD & $0.060$ & $0.071$ & $0.110$ & $0.130$ & $0.200$ & $0.209$  \\
    &&RMSE& $0.0768$ & $0.1077$ & $0.1143$ & $0.1360$ & $0.2241$ & $0.2166$ \\
    &$\widehat{M}_{\hat{Q}}$ & Mean & $5.758$ & $7.803$ & $8.593$ & $8.964$ & $9.214$ & $9.260$ \\
    && SD & $0.002$ & $0.009$ & $0.005$ & $0.009$ & $0.005$ & $0.003$ \\
    &&RMSE& $3.4980$& $4.6000$& $4.7150$ & $4.3920$ & $2.8460$& $2.3840$\\
\hline
 \end{tabular}
 \footnotetext{Note: Estimates are measured in 1000's.}
\label{tablesrm2}
\end{table}

\begin{table}[htbp]
\centering
\caption{Mean, SD \& RMSE of exponential SRM estimators for $F_P=PaI(x_0=10^3, \alpha=2)$ distribution.}

\begin{tabular}{|l|l|l|l|l|l|l|l|l|}
\hline
n & Estimation Methods & Estimated Values & $k=1$ & $k=5$ & $k=10$ & $k=20$ & $k=100$ & $k=200$ \\ [0.5ex]
\hline
     & Theoretical Values & & $2.363$ & $3.984$ & $5.605$ & $7.927$ & $17.725$ & $24.066$ \\
\hline
$30$ & $\widehat{M}_{ML}$ & Mean & $2.469$ & $4.610$ & $6.501$ & $12.653$ & $34.498$ & $53.617$ \\
    && SD & $0.470$ & $1.902$ & $1.638$ & $8.869$ & $32.004$ & $54.988$ \\
    &&RMSE&$0.4818$& $2.0024$ & $1.8670$ & $10.0496$ & $3.64153$ & $55.7764$ \\
    &$\widehat{M}_{PM}$& Mean & $2.280$ & $3.330$ & $5.343$ & $7.561$ & $17.134$ & $24.493$ \\
    && SD & $0.471$ & $0.949$ & $1.986$ & $3.289$ & $9.984$ & $15.827$  \\
    &&RMSE&$0.4783$& $1.1525$& $2.0032$ & $3.3093$ & $10.0015$& $15.8328$\\
    &$\widehat{M}_{EMP}$& Mean & $5.719$ & $5.729$ & $5.746$ & $6.719$ & $11.037$ & $12.701$ \\
    && SD & $0.540$ & $0.526$ & $0.520$ & $0.492$ & $1.961$ & $1.584$ \\
    &&RMSE& $3.3992$& $1.8226$& $0.5388$& $1.3071$& $6.9696$& $11.4749$\\
    &$\widehat{M}_{Prod}$& Mean & $2.539$ & $4.009$ & $5.140$ & $8.002$ & $17.643$ & $24.139$ \\
    && SD & $0.422$ & $0.918$ & $1.102$ & $2.305$ & $8.104$ & $12.100$ \\
    &&RMSE&$0.4572$& $0.9183$& $1.1961$& $2.3062$& $8.1044$& $12.1002$\\
    &$\widehat{M}_{\hat{Q}}$ & Mean & $5.954$ & $8.005$ & $8.978$ & $9.284$ & $9.447$ & $9.615$ \\
    && SD & $0.120$ & $0.188$ & $0.221$ & $0.311$ & $0.135$ & $0.196$ \\
    &&RMSE& $3.5930$& $4.0254$& $3.3802$& $1.3922$& $8.2791$& $14.4523$\\
$100$ & $\widehat{M}_{ML}$ & Mean & $2.671$ & $4.951$ & $7.982$ & $10.993$ & $25.073$ & $35.235$ \\
    && SD & $0.410$ & $0.838$ & $1.856$ & $2.835$ & $8.823$ & $16.445$  \\
    && RMSE & $0.5128$ & $1.2796$ & $3.0158$ & $10.5876$ & $11.4821$ & $19.8793$\\
    &$\widehat{M}_{PM}$& Mean & $2.364$ & $3.578$ & $5.542$ & $7.767$ & $18.523$ & $24.865$ \\
    && SD & $0.275$ & $0.648$ & $1.202$ & $2.317$ & $8.852$ & $11.251$  \\
    &&RMSE& $0.2750$& $0.7647$& $1.2037$& $2.3225$& $8.8879$& $11.2793$\\
    &$\widehat{M}_{EMP}$& Mean & $5.680$ & $5.690$ & $5.699$ & $6.701$ & $11.207$ & $13.642$  \\
    && SD & $0.288$ & $0.288$ & $0.293$ & $0.285$ & $1.317$ & $0.702$  \\
    &&RMSE&$3.3295$& $1.7301$& $0.3077$& $1.2587$& $6.6497$& $10.4476$\\
    &$\widehat{M}_{Prod}$& Mean & $2.431$ & $3.941$ & $5.739$ & $7.810$ & $17.739$ & $24.041$  \\
    && SD & $0.229$ & $0.528$ & $0.927$ & $1.029$ & $4.028$ & $8.030$  \\
    &&RMSE&$0.2389$& $0.5297$& $0.9366$& $1.0356$& $4.0280$& $8.0300$\\
    &$\widehat{M}_{\hat{Q}}$ & Mean & $5.822$ & $7.863$ & $8.714$ & $9.044$ & $9.288$ & $9.352$ \\
    && SD & $0.035$ & $0.038$ & $0.050$ & $0.043$ & $0.044$ & $0.086$ \\
    &&RMSE&$3.4592$& $3.8792$& $3.1094$& $1.1178$& $8.4371$& $14.7143$\\
$500$& $\widehat{M}_{ML}$ & Mean & $2.753$ & $4.821$ & $7.044$ & $10.776$ & $26.621$ & $39.320$  \\
    && SD & $0.206$ & $0.429$ & $0.681$ & $1.584$ & $5.169$ & $8.446$ \\
    &&RMSE&$0.4411$& $0.9405$ & $1.5920$ & $3.2597$ & $10.2887$ & $17.4362$\\
    &$\widehat{M}_{PM}$& Mean & $2.351$ & $3.957$ & $5.678$ & $7.515$ & $17.573$ & $24.858$  \\
    && SD & $0.136$ & $0.338$ & $0.454$ & $1.381$ & $2.571$ & $4.299$  \\
    &&RMSE& $0.1365$& $0.3391$& $0.4598$& $1.4411$& $2.5755$& $4.3713$\\
    &$\widehat{M}_{EMP}$& Mean & $5.654$ & $5.659$ & $5.662$ & $6.665$ & $11.313$ & $13.972$  \\
    && SD & $0.125$ & $0.125$ & $0.128$ & $0.130$ & $0.701$ & $0.147$  \\
    &&RMSE& $3.2913$& $1.8158$& $0.1419$& $1.2685$& $6.4132$& $10.0948$\\
    &$\widehat{M}_{Prod}$& Mean & $2.323$ & $3.908$ & $5.708$ & $7.908$ & $17.588$ & $24.008$  \\
    && SD & $0.121$ & $0.290$ & $0.399$ & $0.945$ & $2.121$ & $3.909$  \\
    &&RMSE&$0.1274$& $0.2998$& $0.4121$& $1.2505$& $2.1254$& $3.9094$\\
    &$\widehat{M}_{\hat{Q}}$ & Mean & $5.769$ & $7.814$ & $8.594$ & $8.973$ & $9.229$ & $9.277$ \\
    && SD & $0.013$ & $0.012$ & $0.008$ & $0.012$ & $0.014$ & $0.016$ \\
    &&RMSE& $3.4060$& $3.8300$& $2.9890$& $1.0461$& $8.4960$& $14.7890$\\
\hline
 \end{tabular}
 \footnotetext{Note: Estimates are measured in 1000's.}
\label{tablesrm3}
\end{table}


\subsection{Dependent case}

To study the effect of dependence we consider the model described in \cite{liang2015}. In this case we avoid the parametric approach as it is difficult to establish the parametric form using the model described below. So we have compared three estimators of exponential SRM namely $\widehat{M}_{EMP}$, $\widehat{M}_{Prod}$ and $\widehat{M}_{\hat{Q}}$. Let, ($X1$,$Y$,$T$,$\delta$) denote a random vector where $X1$ is an $\mathbb{R}$-valued random vector of covariates related with $X$ where $Y=\min(X,\ S)$ and $\delta=I(X\leq S)$. 
Here $\alpha=P(Y\geq T)$. 
Now, in order to obtain an $\alpha$-mixing observed sequence \{$X1_i$,$Y_i$,$T_i$,$\delta_i$\}, the observed data is generated as follows: 
\begin{eqnarray*}
X1_1&=&0.5e_1, \\
X1_i&=&\rho X1_{i-1}+0.5e_i, \\
X_i&=&\sin(\pi X1_i)+\phi_1(1+0.3\cos(\pi X1_i))\epsilon_i, \\
S_i&=&\sin(\pi X1_i)+0.5\phi_2(1+0.3\cos(\pi X1_i))+\phi_3(1+0.3\cos(\pi X1_i))\tilde{\epsilon}_i, \\
Y_i&=&X_i\wedge S_i,\ \text{and}\ \delta_i=I(X_i\leq S_i),
\end{eqnarray*}
where $e_i\sim N(0,1)$, $\epsilon_i\sim N(0,1)$, $\tilde{\epsilon}_i\sim N(0,1)$, and $T_i\sim N(\mu,1)$; everything is conditionally distributed on $Y_i\leq T_i$ and $|\rho|<1$ is some constant, chosen to control the dependence of the observations. It should be noted that the observable $X1_i$'s $\alpha$-mixing property is immediately transferred to the ($X1_i$,$Y_i$,$T_i$,$\delta_i$). Also, note that $$X|_{X1=x}\sim N(\sin(\pi x),\phi_1^2(1+0.3\cos(\pi x))^2).$$

Additionaly, the parameters $\phi_i(i=1,2,3)$, which allow for the control of the percentage of censoring (PC), are provided by
$$PC=P(X_i>W_i|X1_i=x)=1-\Phi\left(\frac{0.5\phi_2}{\sqrt{\phi_1^2+\phi_3^2}}\right).$$
\cite{liang2015} suggested to use $\phi_1=\phi_3=0.3$ then

\begin{equation} 1-\Phi\Big(\frac{5\sqrt{2}\phi_2}{6}\Big)=\left\{ \begin{array}{ll}10\%& \mathrm{when}\quad \phi_2=1.087 \\
15\% & \mathrm{when}\quad \phi_2=0.8796 \\
30\% & \mathrm{when}\quad \phi_2=0.445.\end{array}\right.\end{equation}

\subsubsection{Simulation setting and results}

Next we draw random samples with sample size $n=30,\ 100,\ 500$ and $\rho=0.1$ respectively from the above model. In Table 3 we report the means, SDs and RMSEs of the exponential SRM estimates for truncation rate ($\alpha=30\%$) and PC is equal to $10\%$. In Figure 3 we plot the logarithm of ratio of the RMSEs of different exponential SRM estimators for the dependent case and we can clearly observe that for small values of $k$, $\widehat{M}_{Prod}$ out performs all the estimators. From Table 3 we observe that for all the sample sizes $n$ and for $k=5,\ 10$ the $\widehat{M}_{Prod}$ estimator outperforms all the estimator and when $n\geq100$ the $\widehat{M}_{Prod}$ estimator also outperforms the other estimator for $k=20$. For $n\leq500$ and for all values of $k$ except $k=100$, $\widehat{M}_{\hat{Q}}$ estimator outperforms all the estimators. We also observe that when $n\leq 100$ and $k=1$ the $\widehat{M}_{EMP}$ estimator out performs all the estimators.

\begin{table}[htbp]
\centering
\caption{Mean, SD \& RMSE of exponential SRM estimators for dependent case when $\rho=0.1$, $\alpha=30\%$ and $PC=10\%$.}
\begin{tabular}{|l|l|l|l|l|l|l|l|l|}
\hline
$n$ & Estimation Methods & Estimated Values & $k=1$ & $k=5$ & $k=10$ & $k=20$ & $k=100$ & $k=200$ \\ [0.5ex]
\hline
     & Theoretical Values & & $0.1682$ & $0.6633$ & $0.9202$ & $1.0679$ & $1.1750$ & $1.2099$ \\
\hline
$30$ & $\widehat{M}_{Prod}$ & Mean & $0.1646$ & $0.6958$ & $0.9098$ & $1.1915$ & $1.8261$ & $3.6521$ \\
    && SD & $0.3610$ & $0.3234$ & $0.4228$ & $0.7826$ & $3.9987$ & $7.9975$ \\
    &&RMSE & $0.3610$ & $0.3250$ & $0.4230$ & $0.7923$& $4.0514$ & $8.3621$ \\
    &$\widehat{M}_{EMP}$& Mean & $0.2799$ & $1.0899$ & $1.7942$ & $3.2875$ & $16.2685$ & $32.5531$ \\
    && SD & $0.1700$ & $0.1241$ & $0.0829$ & $0.0953$ & $0.5309$ & $0.9704$ \\
    &&RMSE& $0.2045$ & $0.4443$ & $0.8772$ & $2.2216$ & $15.1028$ & $31.3582$ \\
    &$\widehat{M}_{\hat{Q}}$& Mean & $0.1711$ & $0.2740$ & $0.3155$ & $0.4188$ & $2.0475$ & $4.0083$ \\
    && SD & $0.3645$ & $0.1787$ & $0.1198$ & $0.1550$ & $0.7192$ & $1.5275$ \\
    &&RMSE& $0.3646$& $0.4284$ & $0.6172$ & $0.6674$ & $1.1307$ & $3.1882$ \\
$100$ & $\widehat{M}_{Prod}$ & Mean & $0.1325$ & $0.6895$ & $0.9016$ & $1.0784$ & $1.2189$ & $1.6103$ \\
    && SD & $0.2034$ & $0.1898$ & $0.2482$ & $0.4916$ & $2.5723$ & $5.0307$ \\
    &&RMSE & $0.2064$& $0.1916$& $0.2490$ & $0.4917$ & $2.5727$ & $5.0466$ \\
    &$\widehat{M}_{EMP}$& Mean & $0.2734$ & $1.1257$ & $1.8494$ & $3.3618$ & $16.6698$ & $33.3238$ \\
    && SD & $0.0935$ & $0.0683$ & $0.0740$ & $0.1221$ & $0.5221$ & $0.7908$ \\
    &&RMSE& $0.1407$&$0.4674$& $0.9315$& $2.2972$& $15.5036$& $32.1236$ \\
    &$\widehat{M}_{\hat{Q}}$& Mean & $0.1323$ & $0.2282$ & $0.3217$ & $0.4173$ & $2.0192$ & $4.0680$ \\
    && SD & $0.1937$ & $0.1064$ & $0.0649$ & $0.0873$ & $0.4350$ & $0.8328$ \\
    &&RMSE& $0.1970$& $0.4479$& $0.6027$& $0.6564$& $0.9497$& $2.9770$ \\
$500$& $\widehat{M}_{Prod}$ & Mean & $0.1908$ & $0.6568$ & $0.9109$ & $1.0817$ & $1.1247$ & $2.0493$ \\
    && SD & $0.0894$ & $0.0915$ & $0.1180$ & $0.2348$ & $1.1093$ & $2.2185$ \\
    &&RMSE& $0.0922$& $0.0917$& $0.1183$& $0.2352$& $1.1104$& $2.3720$ \\
    &$\widehat{M}_{EMP}$& Mean & $0.2817$ & $1.1536$ & $1.8951$ & $3.4544$ & $17.0628$ & $34.166$ \\
    && SD & $0.0935$ & $0.0617$ & $0.1201$ & $0.2463$ & $1.0056$ & $2.1880$ \\
    &&RMSE& $0.1473$& $0.4942$& $0.9816$& $2.3992$& $15.9196$& $33.0287$ \\
    &$\widehat{M}_{\hat{Q}}$& Mean & $0.1267$ & $0.2177$ & $0.3219$ & $0.4203$ & $2.0222$ & $4.0378$ \\
    && SD & $0.0889$ & $0.0511$ & $0.0299$ & $0.0380$ & $0.1975$ & $0.3810$ \\
    &&RMSE& $0.0980$& $0.4486$& $0.5997$ & $0.6487$& $0.3666$ & $2.8535$ \\
\hline
 \end{tabular}
\label{tablesrm5}
\end{table}

\subsection{Findings from the simulation study}

From the simulations we find that our proposed estimator $\widehat{M}_{Prod}$ performs very well for small sample size $n\leq100$ and small values of $k$ for $F_E$ distribution and for large sample size we can consider $\widehat{M}_{ML}$ estimator. For $F_P$ distribution our proposed estimator $\widehat{M}_{Prod}$ outperforms other estimators for large sample size $100\leq n$ and for almost all values of $k$ and for small sample it works only when $k=1$ and $5$. We also observe that for $n=30$, $\widehat{M}_{EMP}$ is better for some values of $k$ but far worse for small values of $k$ in the i.i.d case. In the dependent case $\widehat{M}_{Prod}$ is clearly better than all other methods for all sample sizes and small values of $k$ i.e. for 5 and 10 in terms of RMSE. When the sample size is large $\widehat{M}_{Prod}$ out performs for all values of $k$ except for $k=100$.

\subsection{Coverage probabilities}
This section conducts a simulation study, where we estimate the coverage probabilities of confidence intervals that capture the population parameter of interest. We have considered the $F_E$ and $F_P$ which are defined in section (\ref{Ex}). For each distribution $F_E$ and $F_P$, we created 10,000 confidence intervals and each $\widehat{M}_{Prod}^*$ has been obtained from 1000 bootstrap replicates. For nominal confidence level $90\%$ we have obtained coverage probabilities with $n=30$, $100$, $500$ and $k=1$, $5$, $10$, $20$, $100$ and $200$. The results are reported in Table \ref{tablecov}. From Table \ref{tablecov} we observe that for higher values of $k$ the coverage probabilities are too low. However, for small values of $k$ the coverage probabilities are close to the nominal confidence level $90\%$.

\begin{table}[htbp]
\centering
\caption{Simulated coverage probabilities of the bootstrap confidence intervals for $F_E$ and $F_P$ distribution.}

\begin{tabular}{|l|l|l|l|l|}
\hline
Distribution & $k$ & 30 & 100 & 500  \\ [0.5ex]
\hline
 &&\multicolumn{3}{|c|}{$n$} \\
 \hline
$F_E$ & 1 & 0.872 & 0.89 & 0.92  \\
    &5& 0.86 & 0.871 & 0.911 \\
    &10& 0.85 & 0.86 & 0.901 \\
    &20& 0.82 & 0.84 & 0.891  \\
    &100& 0.79 & 0.83 & 0.89 \\
    &200& 0.77 & 0.80 & 0.85 \\
$F_P$ & 1 & 0.88 & 0.89 & 0.91 \\
    &5& 0.85 & 0.872 & 0.89  \\
    &10&0.821 & 0.832 & 0.876 \\
    &20&0.8 & 0.82 & 0.85 \\
    &100& 0.75 & 0.777 & 0.80 \\
    &200& 0.73 & 0.74 & 0.78 \\
\hline
 \end{tabular}
\label{tablecov}
\end{table}

\section{Data analysis}

To illustrate our work, we use the thoroughly researched Norwegian fire claims data \cite{upretee2020} and the French marine losses data available in the ``CASdatasets'' package in R software. The Norwegian fire claims data show the overall amount of harm caused by fires in Norway from 1972 to 1992. It's worth noting that only damages in excess of $500,000$ Norwegian krones (NOK) are covered. It's also unclear whether the claims were adjusted for inflation. The French marine losses dataset comprises of $1,274$ marine losses between $2003$ and $2006$. The damages lying between between $0.018$ and $31904.2$ Euros are covered. Table \ref{tablesum} summarizes the first data set over the last $12$ years, from $1981$ to $1992$. Table \ref{SS} summarizes the second data set. Table \ref{tablesum} and \ref{SS} shows that these data sets contain all of the standard characteristics of insurance claims. Specifically, majority of frequently occurring claims are rather small, but as severity increases, claim frequency decreases and a small fraction of extremely big claims occur.\

The first data set is analyzed for the final $12$ years, $1981$--$1992$, to compare our findings with that of \cite{upretee2020} and \cite{brazauskas2016}. Due to the fact that no information is provided below $500,000$ and there is no policy limitation, the random variable that generated the data is left truncated at $d=500,000$ but is not censored. Similarly, the random variable that generates the second data set is LTRC because no information is provided below $d=0.018$ and there exists policy limitation i.e. $u=31904.2$. We calculate the exponential SRM using our proposed estimator $\widehat{M}_{Prod}$. In Table \ref{Q} and \ref{T} we report the point estimates and $90\%$ confidence intervals (CIs) of exponential SRM using the PL estimator for the mentioned data sets. The corresponding CIs are constructed using the asymptotic distribution given in Theorem \ref{theo2}. The CIs are constructed using the sample size of the claims from $1981$--$1992$ for the first data set and from $2003$--$2006$ for the second data set. We run $1000$ bootstrap samples. In each iteration, a random sample of size equal to the sample size of the data corresponding to each year is selected with replacement from the data. Using each sample, the lower and upper limits of the confidence intervals are calculated.\

If we observe the exponential SRM values in Table \ref{Q} and \ref{T}, several patterns can be observed. We notice that the riskiness of Norwegian fire claims does not show any discernible trend over time, neither higher nor decrease, as seen in \cite{brazauskas2016}. Similar patterns are also observed for the French marine losses. From Table \ref{Q} we observe that CIs are narrow for small values of $k$ compared to the CIs for higher values of $k$ for Norwegian fire claims where we see that the sample size lies between $400$--$900$ from $1981$--$1992$. In case of French Marine losses we observe similar patterns, CIs are much wider for higher values of $k$ compared to the CIs for small values of $k$ where we see that the sample size lies between $90$--$400$ (see table \ref{T}). This is because an SRM estimator predicated on a high value of $k$ (when $k\geq100$) is a weighted average of observations that puts a lot of weight on a small subset of observation, and therefore operates with a smaller effective sample size. \

\begin{table}[htbp]
\centering
\caption{Summary statistics for the Norwegian Fire Claims ($1981$-$1992$) data.}

 \begin{tabular}{|l|l|l|l|l|l|l|l|l|l|l|l|l|}
\hline
 $\shortstack{Claim severity\\ (in 1000,000's)}$ & $1981$ & $1982$ & $1983$ & $1984$ & $1985$ & $1986$ & $1987$ & $1988$ & $1989$ & $1990$ & $1991$ & $1992$  \\ [0.5ex]
 \hline
[$0.5\ 1.0$) & $54.5$ & $54.7$ & $53.8$ & $48.3$ & $49.6$ & $50.7$ & $41.9$ & $41.2$ & $41.9$ & $40.4$ & $43.3$ & $45.5$  \\
\hline
[$1.0\ 2.0$) & $23.5$ & $23.8$ & $27.8$ & $29.8$ & $29.2$ & $31.2$ & $34.7$ & $32.8$ & $33.3$ & $39.0$ & $34.0$ & $31.4$  \\
\hline
[$2.0\ 5.0$) & $13.5$ & $14.3$ & $12.0$ & $15.4$ & $14.0$ & $11.9$ & $17.1$ & $16.9$ & $18.0$ & $16.2$ & $17.9$ & $16.4$\\
\hline
[$5.0\ 10.0$) & $4.4$ & $4.7$ & $3.4$ & $4.8$ & $4.0$ & $3.2$ & $4.2$ & $5.2$ & $3.9$ & $1.9$ & $3.2$ & $4.1$ \\
\hline
[$10.0\ 20.0$) & $2.3$ & $2.1$ & $2.2$ & $1.1$ & $1.6$ & $1.7$ & $1.2$ & $1.9$ & $1.9$ & $1.4$ & $1.3$ & $1.6$ \\
\hline
[$20.0\ \infty$) & $1.6$ & $0.5$ & $0.7$ & $0.5$ & $1.6$ & $1.2$ & $1.0$ & $1.9$ & $1.0$ & $1.0$ & $0.3$ & $1.0$ \\
\hline
Top three claims & $43$ &$19$ &$22$ &$22$ &$60$ &$87$ &$35$ &$84$ &$45$ &$26$ & $17$&$45$ \\
                &$62$&$20$&$30$&$56$&$70$&$98$&$38$&$151$&$86$&$41$& $35$& $50$ \\
                &$78$&$23$&$51$&$106$&$135$&$188$&$45$&$465$&$145$&$79$& $50$& $102$ \\
\hline
Sample size & $429$ & $428$ & $407$ & $557$ & $607$ & $647$ & $767$ & $827$ & $718$ & $628$ & $624$ & $615$ \\ [1ex]
\hline
 \end{tabular}
\footnotetext{Note: Relative frequencies are in $\%$.}
\label{tablesum}
\end{table}

\begin{sidewaystable}[htbp]
\centering
\caption{Point and interval estimates of exponential SRM for Norwegian Fire Claims data using $\widehat{M}_{Prod}$ estimator (measured in 1000 millions NOK).}
 \begin{tabular}{|l|l|l|l|l|l|l|}
\hline
Year & $1$ & $5$ & $10$ & $20$ & $100$ & $200$  \\ [0.5ex]
 \hline
 &\multicolumn{6}{|c|}{$k$}\\
 \hline
$1981$ & $33.533$ & $57.652$ & $65.606$ & $67.557$ & $80.167$ & $114.396$  \\
 & [31.265 35.802] &[53.332 61.179] & [61.095 70.117] &[62.908 72.205] & [74.651 85.683] & [106.523 122.268] \\
$1982$ & $11.391$ & $18.821$ & $21.348$ & $21.928$ & $26.019$ & $37.128$  \\
& [11.143 11.639] & [18.394 19.248] & [20.857 21.840] &[21.422 22.435] & [25.418 26.620] & [36.271 37.986] \\
$1983$ & $20.392$ & $34.522$ & $39.469$ & $40.623$ & $48.206$ & $68.788$   \\
& [17.634 23.149] & [29.744 39.301] &[33.970 44.968] &[34.955 46.291] & [41.479 54.932]& [59.189 78.386] \\
$1984$ & $42.496$ & $72.997$ & $83.850$ & $86.409$ & $102.542$ & $146.323$  \\
 &[40.741 44.251] & [69.944 76.050]  &[80.331 87.369] &[82.779 900.38] &[98.234 106.849] &[140.177 152.469] \\
$1985$ & $52.715$&$90.647$&$104.117$&$107.283$&$127.312$&$181.670$ \\
 &[48.957 56.473] & [84.117 97.176] & [96.595 111.640] & [99.526 115.040]&[118.107 136.518] &[168.534 194.806] \\
 $1986$ & $70.876$ & $122.357$ & $140.733$ & $145.066$ & $172.152$ & 245.655  \\
 &[62.276 79.476] & [107.369 137.345] &[123.439 158.026] & [127.225 162.908] &[150.978 193.326] & [215.441 275.869] \\
 $1987$ & $21.423$ & $36.201$ & $41.372$ & $42.580$ & $50.527$ & 72.101   \\
 &[21.248 21.598] & [34.305 36.493] &[41.041 41.703] & [40.330 42.919] &[50.124 50.930] & [68.290 72.676] \\
 $1988$ & $162.475$ & $279.812$ & $322.246$ & $332.279$ & $394.325$ & $526.867$  \\
 &[143.167 181.782] & [246.934 312.693] &[284.330 360.163] & [293.168 371.390] &[347.910 440.740] & [460.635 593.099] \\
 $1989$ & $61.432$ & $103.962$ & $120.091$ & $126.341$ & $159.363$ & 195.368 \\
 &[54.832 68.032] & [93.433 114.491] &[107.568 132.614] & [113.570 139.110] &[143.157 175.569] & [175.232 215.504]\\
 $1990$ & $38.130$ & $60.022$ & $67.923$ & $70.541$ & $86.903$ & 117.437   \\
 &[35.561 40.699] & [56.096 63.948] &[63.025 72.821] & [65.542 75.540] &[81.090 92.717] & [108.536 126.338]\\
 $1991$ & $18.023$ & $31.090$ & $36.230$ & $39.999$ & $44.921$ & 66.236   \\
 &[15.909 20.137] & [28.100 34.081] &[32.335 40.127] & [35.998 44.000] &[40.031 49.811] & [60.236 72.235] \\
 $1992$ & $40.904$ & $69.803$ & $81.990$ & $85.334$ & $97.892$ & 141.266  \\
 &[39.301 42.507] & [66.814 72.792] &[78.570 85.410] & [81.722 88.946] &[93.440 102.344] & [135.281 147.251] \\[1ex]
      \hline
 \end{tabular}
\label{Q}
\end{sidewaystable}

\begin{table}[htbp]
\centering
\caption{Summary statistics for French marine losses ($2003$-$2006$) data.}\label{SS}

 \begin{tabular}{|l|l|l|l|l|}
\hline
 $\shortstack{Claim \\ (in 100)}$ & $2003$ & $2004$ & $2005$ & $2006$  \\ [0.5ex]
 \hline
[$0.00018\ 0.1$) & $72.81$ & $72.04$ & $72.15$ & $79.12$    \\
\hline
[$0.1\ 1$) & $22.36$ & $24.19$ & $25.63$ & $20.88$   \\
\hline
[$1\ 319.042$) & $4.83$ & $3.76$ & $2.22$ & $0$  \\
\hline
Top three claims & $61$ &$11$ &$1.574$ &$0.413$  \\
                &$12$&$13$&$1.898$&$0.503$ \\
                &$19$&$37$&$2.117$&$0.75$ \\
\hline
Sample size & $331$ & $372$ & $316$ & $91$  \\ [1ex]
\hline
 \end{tabular}
\footnotetext{Note: Relative frequencies are in $\%$.}
\end{table}

\begin{table}[htbp]
\centering
\caption{Point and interval estimates of exponential SRM for French marine losses using $\widehat{M}_{Prod}$ estimator (measured in Euros).}
 \begin{tabular}{|c|c|c|c|c|}
\hline
$k$ & $2003$ & $2004$ & $2005$ & $2006$   \\ [0.5ex]
 \hline
$1$ & $770.063$ & $1188.592$ & $89.363$ & $20.821$  \\
 & [746.2528 793.8738] &[978.254 1398.93] & [87.1068 91.6198] &[20.1753 21.468]  \\
$5$ & $1375.489$ & $2142.582$ & $161.168$ & $44.071$  \\
& [1333.401 1417.577] & [1675.646 2529.953] & [156.8913 165.445] &[43.074 45.069]  \\
$10$ & $1609.593$ & $2519.635$ & $190.398$ & $58.243$   \\
& [1560.62 1658.566] & [1969.058 2980.547] &[187.2089 195.588] &[57.2422 59.244]  \\
$20$ & $1669.925$ & $2619.16$ & $198.540$ & $65.348$  \\
 &[1619.238 1720.612] & [2137.746 3100.574]  &[187.7554 204.010] &[64.4723 66.223]  \\
$100$ & $1982.474$ & $3109.755$ & $235.801$ & $78.701$  \\
 &[1922.312 2042.636] & [2537.982 3681.528] & [229.3007 242.302] & [77.7235 79.678] \\
 $200$ & $2828.916$ & $4437.504$ & $336.480$ & $112.303$ \\
 &[2440.356 2914.765] & [3621.606 5253.402]  &[318.1976 345.756] & [110.9084 113.698]  \\[1ex]
      \hline
 \end{tabular}
%
\label{T}
\end{table}


\section{Conclusion remarks}

As stated in the introduction, the application of risk measures in various insurance problems has been the subject of much discussion in the literature. According to \cite{pichler2015}, the premium principles in an actuarial setting are equivalent to risk measures in financial mathematics. Traditionally, premiums are calculated using the coherent risk measures. In this paper we use the more general class of risk measure known as SRM because of its coherent nature and the advantage of relating the risk measure to the user's risk aversion. The literature makes it abundantly evident that LTRC data are frequently encountered when working with insurance data. For instance, standard insurance deductibles are a typical illustration of left truncation, while policy limitations are a regular example of right censoring. Since the real sample size is never known due to truncation, neglecting this could lead to an incorrect estimation of SRM, which could then have an impact on how premiums are implemented. It is therefore important to see how efficiently we can estimate SRM with respect to LTRC data and validate these estimates.\

We propose a nonparametric estimator of SRM $\widehat{M}_{Prod}$, which is based on the PL estimator, since the simplest nonparametric technique is inefficient due to scarcity of data. We have shown that our estimator is asymptotically normal for i.i.d case and in order to observe the higher-order corrections upon the normal approximations we derived the Edgeworth expansion of our estimator. Next we use the Efron's bootstrap method to approximate the distribution of $\widehat{M}_{Prod}$ and showed that the distribution approximates to its true distribution with an error term $o(n^{-1/2})$. This means that the coverage error for the bootstrap confidence interval is $o(n^{-1/2})$.\

In order to estimate SRM, we used the exponential risk aversion function. Based on the simulation study, we found that the choice of $k$ plays a vital role. We have compared our proposed estimator $\widehat{M}_{Prod}$ with the existing parametric and nonparametric estimators such as $\widehat{M}_{EMP}$, $\widehat{M}_{\hat{Q}}$, $\widehat{M}_{ML}$ and $\widehat{M}_{PM}$. Our study reveals that $\widehat{M}_{Prod}$ is a suitable estimator of SRM when we are dealing with LTRC data. We observe that for small values of $k$ and for small samples, $\widehat{M}_{Prod}$ estimator is the best choice when the simulated data is i.i.d and if the simulated data is dependent, $\widehat{M}_{Prod}$ is the best choice for small $k$'s and all sample sizes. We also observe that the coverage probabilities obtained through simulations are close to the nominal confidence level $90\%$ when $k$'s are small.\

We estimate the exponential SRM of two data sets viz: Norwegian fire claims and French marine losses. The Norwegian fire claims data is left truncated and is analyzed from $1981$--$1992$. The French marine losses data is LTRC and comprises data from $2003$--$2006$. We observe that the riskiness of both the data sets does not exhibit any obvious trend, neither upward nor downward. A similar pattern can also be seen in the results of \cite{upretee2020}, where Norwegian fire claims data was used. We even observe that the CIs obtained for both the data sets are wider for higher values of $k$ compared to the lower values of $k$. These observations appears to be true which matches with our simulation study where we see that for small values of $k$ our estimator outperforms other estimators and even the coverage probabilities are close to the nominal confidence level $90\%$ when $k$'s are small.\ 

\begin{figure}[htbp]
    \begin{center}
  \includegraphics[width=140mm]{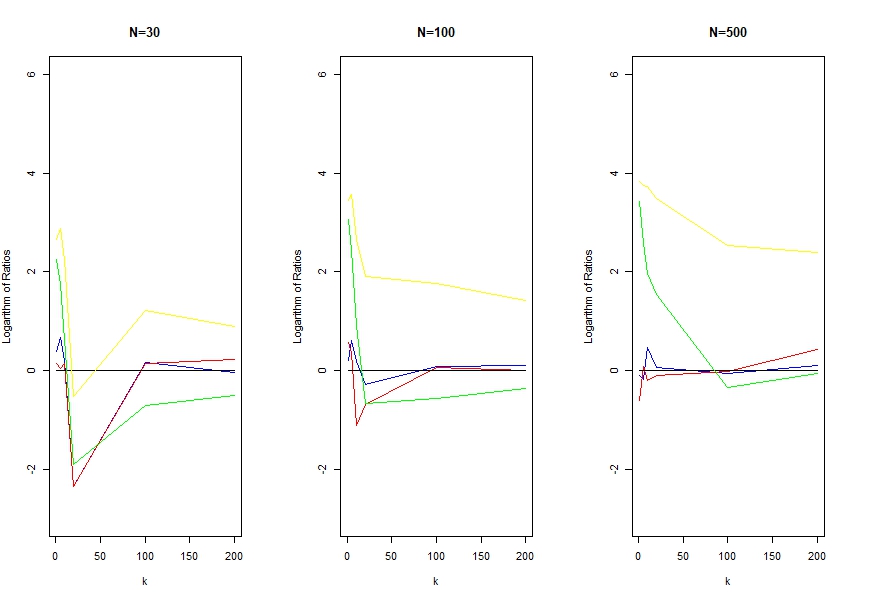}\\
  \caption{Logarithm of ratio of the RMSEs for $F_E$ (black for $\widehat{M}_{Prod}$, blue for $\widehat{M}_{PM}$, red for $\widehat{M}_{ML}$, green for $\widehat{M}_{EMP}$, yellow for $\widehat{M}_{\hat{Q}}$).}
  \end{center}
  \label{fig1}
\end{figure}

\begin{figure}[h]
    \begin{center}
  \includegraphics[width=135mm]{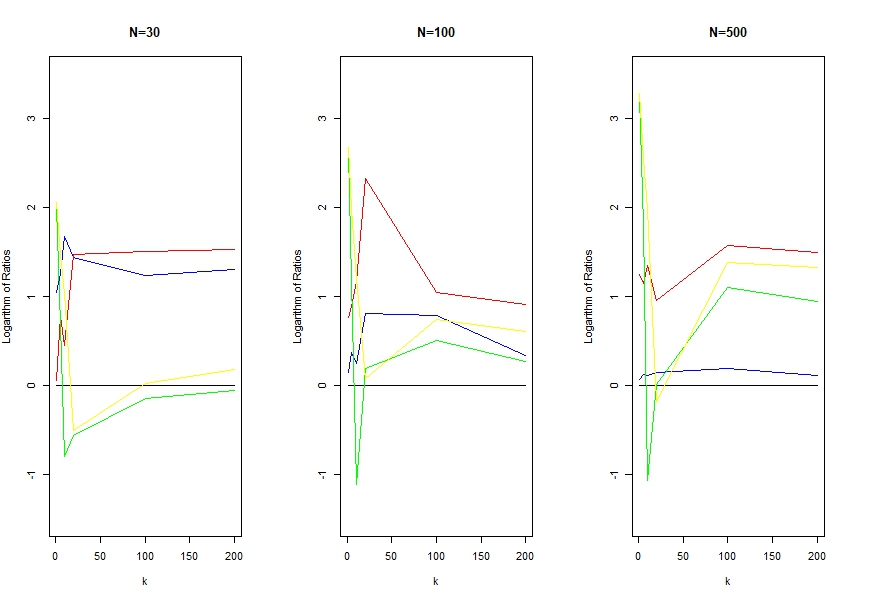}\\
  \end{center}
  \caption{Logarithm of ratio of the RMSEs for $F_P$ (black for $\widehat{M}_{Prod}$, blue for $\widehat{M}_{PM}$, red for $\widehat{M}_{ML}$, green for $\widehat{M}_{EMP}$, yellow for $\widehat{M}_{\hat{Q}}$).}
  \label{fig2}
\end{figure}

\begin{figure}[htbp]
    \begin{center}
  \includegraphics[width=140mm]{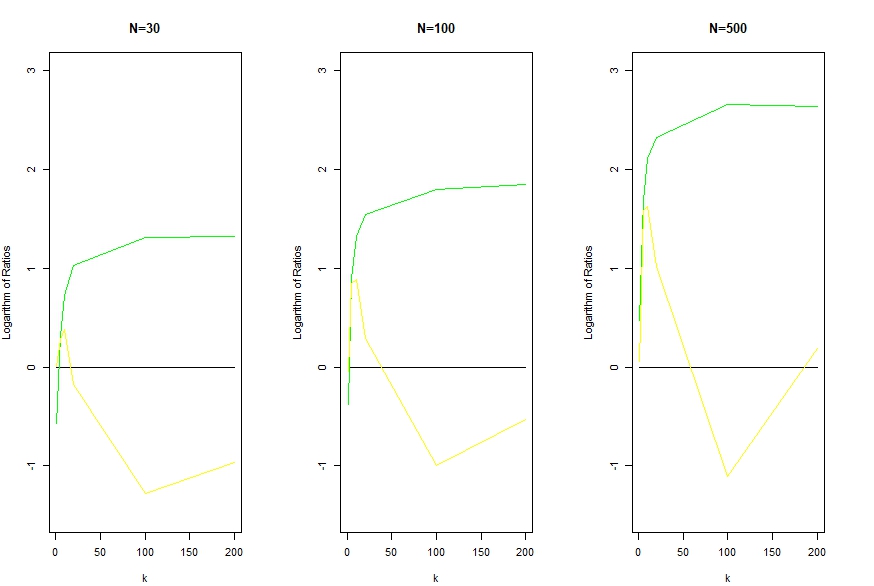}\\
  \end{center}
  \caption{Logarithm of ratio of the RMSEs for dependent case (black for $\widehat{M}_{Prod}$, green for $\widehat{M}_{EMP}$, yellow for $\widehat{M}_{\hat{Q}}$).}
  \label{fig3}
  \end{figure}

\begin{appendices}

\section{Some Definitions}

\begin{definition}\citep{del02}  Let $\psi$ denote the real valued random variables on a probability space ($\Omega,\ \mathcal{F},\ \mathbb{P}$). A coherent risk measure is a function $\rho:\psi\rightarrow\mathbb{R}$ satisfying the following properties:
\begin{enumerate}
\item $X\geq0\Rightarrow\rho(X)\leq0$.
\item $X\geq Y\Rightarrow\rho(X)\leq\rho(Y),\  X,Y\in \psi.$
\item $\rho(\lambda X)=\lambda\rho(X)$, $\forall\lambda\geq0,\ X\in \psi.$
\item $\rho(X+k)=\rho(X)-k$, $\forall k\in\mathbb{R},\ X\in \psi.$
\item $\rho(X+Y)\leq\rho(X)+\rho(Y)$, $\forall X,\ Y \in \psi$.
\end{enumerate}
\end{definition}

\begin{definition}
Let $X\in\psi$ and $F$ be the distribution function of $X$ then $Q_p=\inf\{x:F(x)\geq p\},\ 0<p<1$ is the quantile function.
\end{definition}

\begin{definition} \citep{gzyland08} An element $\phi\in\mathcal{L}^1([0,1])$ is called an admissible risk spectrum if
\begin{enumerate}
\item $\phi\geq0$
\item $\int_0^1|\phi(t)|dt=1$
\item $\phi$ is non-decreasing.
\end{enumerate}
\end{definition}

\section{Some Theorems and Lemmas used in the proof of results discussed in section 3}

\begin{theorem}\label{theo1}\citep{zhou1996} Suppose that assumptions 1 and 2 are satisfied. Then we have uniformly in $a_W\leq x\leq b<b_W$,
\begin{align*}\widehat{F}(x)-F(x)=(1-F(x))L_n(x)+R_{n2}(x) \end{align*} with
\begin{align}\label{con1} \sup_{a_W\leq x\leq b}|R_{n2}(x)|=O(n^{-1}\log\log n)\ a.s.  \end{align}
where $$L_n(x)=\int_{a_W}^x\frac{d(W_{n}^*(y)-W^*(y))}{C(y)}-\int_{a_W}^x\frac{C_n(y)-C(y)}{C^2(y)}dW^*(y).$$
\end{theorem}

\noindent The next result follows from \cite{Gurler1993} that if $a_W<F^{-1}(t_1)\leq F^{-1}(t_2)<b_W$, $F$ has a continuous and positive density $f$ on [$F^{-1}(t_1)-\beta$, $F^{-1}(t_2)+\beta$] for some $\beta>0$ then as $n\rightarrow\infty$,
\begin{align}\label{E1} \sup_{t_1\leq t\leq t_2}\Big|\widehat{F}^{-1}(t)-F^{-1}(t)-\frac{t-\hat{F}(F^{-1}(t))}{f(F^{-1}(t))}\Big|=o_p(n^{-1/2})  \end{align}

\noindent Next theorem, gives the $U$-statistic representation for the PL estimator $\widehat{F}$ with sufficiently small remainder term.
\begin{theorem}\label{theo6.4}\citep{wang06}
Suppose $F$ is a continuous df and $a_G<a_W$. Then, for any fixed $x$ such that $a_W\leq x\leq b<b_W$, we have
\begin{equation*}
\widehat{F}(x)-F(x)=(1-F(x))U^{(n)}-\frac{(1-F(x))\sigma_0^2}{2n}+\alpha_n,\end{equation*}
with $P(\sqrt{n}|\alpha_n|>cn^{-1/2}\log^{(-1)}{n})=o(n^{-1/2})$.
\end{theorem}

\begin{lemma}\label{lem3.5}\citep{wang06}
Let $\sigma_n^2$ be the variance of $U^{(n)}$. Then, under the conditions of Theorem \ref{theo6.4}, for any $a_W\leq x\leq b<b_W$, we have
\begin{equation*}
\sup_y|P(\sqrt{n}\sigma_n^{-1}U^{(n)}\leq y)-E_{n0}(y)|=o(n^{-1/2}),
\end{equation*}
where
\begin{equation*}E_{n0}(y)=\Phi(y)-\frac{\kappa_3}{6}n^{-1/2}\phi_0(y)(y^2-1),\end{equation*}
\begin{equation*}\kappa_3=\frac{1}{\sigma_0^3}\Big(-\frac{15}{2}\sigma_0^4+\int_{a_W}^x\frac{dW^*(t)}{C^3(t)}\Big).  \end{equation*}
\end{lemma}


\begin{lemma}\label{lem3.6}\citep{wang06} For any bounded function $H(y)$, there exists a constant $\alpha$ such that for any random variables $X^0$, $Y^0$, and constant $a>0$
\begin{equation*}\sup_y|P(X^0+Y^0\leq y)-H(y)|\leq\sup_y|P(X^0\leq y)-H(y)|+\alpha a+P(|Y^0|\geq a).   \end{equation*}
\end{lemma}

\section{Proofs of results in section 3}

\noindent\textbf{Proof of Theorem \ref{theo2}:} \begin{eqnarray}\label{E2}\widehat{M}_{Prod}-M_\phi&=&\int_0^1(\widehat{F}^{-1}(u)-F^{-1}(u))\phi(u)du. \end{eqnarray}
Let $a_W<F^{-1}(u_1)\leq F^{-1}(u_2)<b_W$ and using equation (\ref{E1}) and theorem \ref{theo1} we have \footnotesize{\begin{eqnarray*}\widehat{F}^{-1}(u)-F^{-1}(u)&=&-\frac{\widehat{F}(F^{-1}(u))-F(F^{-1}(u))}{f(F^{-1}(u))} \\
&=&\frac{-1}{f(F^{-1}(u))}\Big[(1-F(F^{-1}(u)))\Big(\int_{a_W}^{F^{-1}(u)}\frac{d(W_{n}^*(y)-W^*(y))}{C(y)}-\int_{a_W}^{F^{-1}(u)}\frac{C_n(y)-C(y)}{C^2(y)}dW^*(y)\Big)\Big] \\
&=&\frac{-1}{f(F^{-1}(u))}\Big[(1-u)\Big(\frac{1}{n}\sum_{i=1}^n\frac{I(Y_i\leq F^{-1}(u),\delta_i=1)}{C(Y_i)}-\int_{a_W}^{F^{-1}(u)}\frac{1/n\sum_{i=1}^nI(T_i\leq y\leq Y_i)}{C^2(y)}dW^*(y)\Big)\Big] \\
&=&\frac{-(1-u)}{nf(F^{-1}(u))}\sum_{i=1}^n\Big[\frac{I(Y_i\leq F^{-1}(u),\delta_i=1)}{C(Y_i)}-\int_{a_W}^{F^{-1}(u)}\frac{I(T_i\leq y\leq Y_i)}{C^2(y)}dW^*(y)\Big]
 \end{eqnarray*}}

Now we can write
\begin{eqnarray*}\widehat{M}_{Prod}-M_\phi&=&\frac{1}{n}\sum_{i=1}^n\int_0^1\frac{-(1-u)}{f(F^{-1}(u))}\Big[\frac{I(Y_i\leq F^{-1}(u),\delta_i=1)}{C(Y_i)}-\int_{a_W}^{F^{-1}(u)}\frac{I(T_i\leq y\leq Y_i)}{C^2(y)}dW^*(y)\Big]\phi(u)du
\end{eqnarray*}

Let \begin{align*}T_i=\int_0^1\frac{-(1-u)}{f(F^{-1}(u))}\Big[\frac{I(Y_i\leq F^{-1}(u),\delta_i=1)}{C(Y_i)}-\int_{a_W}^{F^{-1}(u)}\frac{I(T_i\leq y\leq Y_i)}{C^2(y)}dW^*(y)\Big]\phi(u)du.  \end{align*}

\begin{eqnarray*}E(T_i)&=&E\Big[\int_0^1\frac{-(1-u)}{f(F^{-1}(u))}\Big[\frac{I(Y_i\leq F^{-1}(u),\delta_i=1)}{C(Y_i)}-\int_{a_W}^{F^{-1}(u)}\frac{I(T_i\leq y\leq Y_i)}{C^2(y)}dW^*(y)\Big]\phi(u)du\Big] \\
&=&\int_0^1\frac{-(1-u)}{f(F^{-1}(u))}E\Big[\frac{I(Y_i\leq F^{-1}(u),\delta_i=1)}{C(Y_i)}-\int_{a_W}^{F^{-1}(u)}\frac{I(T_i\leq y\leq Y_i)}{C^2(y)}dW^*(y)\Big]\phi(u)du \\
&=&\int_0^1\frac{-(1-u)}{f(F^{-1}(u))}\Big[\frac{P(T\leq F^{-1}(u),\delta=1|T\leq Y)}{C(Y)}-\int_{a_W}^{F^{-1}(u)}E\Big(\frac{I(T_i\leq y\leq Y_i)}{C^2(y)}\Big)dW^*(y)\Big]\phi(u)du \\
&=&\int_0^1\frac{-(1-u)}{f(F^{-1}(u))}\Big[P(T\leq F^{-1}(u),\delta=1|T\leq Y)-\int_{a_W}^{F^{-1}(u)}\frac{P(T\leq y\leq Y|T\leq Y)}{C^2(y)}dW^*(y)\Big]\phi(u)du \\
&=&\int_0^1\frac{-(1-u)}{f(F^{-1}(u))}\Big[P(T\leq F^{-1}(u),\delta=1|T\leq Y)-\int_{a_W}^{F^{-1}(u)}\frac{P(T\leq y\leq Y|T\leq Y)}{C^2(y)}dW^*(y)\Big]\phi(u)du \\
&=&\int_0^1\frac{-(1-u)}{f(F^{-1}(u))}\Big[P(T\leq F^{-1}(u),\delta=1|T\leq Y)-\int_{a_W}^{F^{-1}(u)}\frac{dW^*(y)}{C(y)}\Big]\phi(u)du \\
&=&\int_0^1\frac{-(1-u)}{f(F^{-1}(u))}\Big[P(T\leq F^{-1}(u),\delta=1|T\leq Y)-\frac{W^*(F^{-1}(u))}{C(Y)}\Big]\phi(u)du \\
&=&\int_0^1\frac{-(1-u)}{f(F^{-1}(u))}[P(T\leq F^{-1}(u),\delta=1|T\leq Y)-P(T\leq F^{-1}(u),\delta=1|T\leq Y)]\phi(u)du \\
&=&0
\end{eqnarray*}

\noindent Therefore, \begin{equation*}E\Big[\frac{1}{n}\sum_{i=1}^nT_i\Big]=0.  \end{equation*}

and \begin{equation*}\sigma^2=V\Big[\frac{1}{n}\sum_{i=1}^nT_i\Big]=\int_0^1\int_0^1\frac{(1-u)(1-v)}{f(F^{-1}(u))f(F^{-1}(v))}\Big[\int_{a_W}^{F^{-1}(u)}\frac{dW^*(y)}{C^2(y)}\int_{a_W}^{F^{-1}(v)}\frac{dW^*(y)}{C^2(y)}-uv\Big]\phi(u)\phi(v)dudv.  \end{equation*}
\

\noindent Now an application of Liapunov's form of the Central Limit Theorem we get
\begin{equation*}\sqrt{n}(Prod-\mu)\xrightarrow{\text{d}} N(0,\sigma^2)\ as\ n\rightarrow\infty.\end{equation*}

\noindent\textbf{Proof of Theorem \ref{theo6.5}:} We can write \begin{eqnarray*} \widehat{M}_{Prod}-M_\phi&=&\int_0^1(\widehat{F}^{-1}(u)-F^{-1}(u))\phi(u)du. \end{eqnarray*}
Let, $a_W< F^{-1}(u_1)\leq F^{-1}(u_2)<b_W$ and using equation (\ref{E1}) and theorem \ref{theo6.4} we have
\small{\begin{eqnarray*}\widehat{F}^{-1}(u)-F^{-1}(u)&=&-\frac{\widehat{F}(F^{-1}(u))-F(F^{-1}(u))}{f(F^{-1}(u))} \\
&=&\frac{-1}{f(F^{-1}(u))}\Big[(1-F(F^{-1}(u)))\Big\{\frac{1}{n^2}\sum_{i<j}h_1(V_i;\ V_j;\ F^{-1}(u))-\frac{1}{2n}\int_{a_W}^{F^{-1}(u)}\frac{dW^*(x)}{C^2(x)}\Big\}\Big]+\alpha_n \\
\end{eqnarray*}}
Now,
\small{\begin{eqnarray*}\int_0^1(\widehat{F}^{-1}(u)-F^{-1}(u))\phi(u)du&=&\int_0^1\frac{-1}{f(F^{-1}(u))}\Big[(1-u)\Big\{\frac{1}{n^2}\sum_{i<j}h_1(V_i;V_j;F^{-1}(u))-\frac{1}{2n}\int_{a_W}^{F^{-1}(u)}\frac{dW^*(x)}{C^2(x)}\Big\}\Big]\\
&&\phi(u)du+\alpha_n. \\
\end{eqnarray*}}

\noindent\textbf{Proof of Theorem \ref{theo3.8}:} \begin{eqnarray*} \widehat{M}_{Prod}-M_\phi&=&\int_0^1(\widehat{F}^{-1}(u)-F^{-1}(u))\phi(u)du. \end{eqnarray*}
Let, $a_W< F^{-1}(u_1)\leq F^{-1}(u_2)<b_W$ and using equation (\ref{E1}) and theorem \ref{theo6.4} we have
\small{\begin{eqnarray*}\widehat{F}^{-1}(u)-F^{-1}(u)&=&-\frac{\widehat{F}(F^{-1}(u))-F(F^{-1}(u))}{f(F^{-1}(u))} \\
&=&\frac{-1}{f(F^{-1}(u))}\Big[(1-F(F^{-1}(u)))\Big\{\frac{1}{n^2}\sum_{i<j}h_1(V_i;\ V_j;\ F^{-1}(u))-\frac{1}{2n}\int_{a_W}^{F^{-1}(u)}\frac{dW^*(x)}{C^2(x)}\Big\}\Big]\\
\end{eqnarray*}}
Now,
\small{\begin{eqnarray*}\int_0^1(\widehat{F}^{-1}(u)-F^{-1}(u))\phi(u)du&=&\int_0^1\frac{-1}{f(F^{-1}(u))}\Big[(1-u)\Big\{\frac{1}{n^2}\sum_{i<j}h_1(V_i;V_j;F^{-1}(u))-\frac{1}{2n}\int_{a_W}^{F^{-1}(u)}\frac{dW^*(x)}{C^2(x)}\Big\}\Big]\\
&&\phi(u)du \\
&=&S_1+S_2 \\
\end{eqnarray*}}
where
\begin{equation*}
S_1=-\int_0^1\frac{(1-u)}{n^2f(F^{-1}(u))}\sum_{i<j}h_1(V_i;\ V_j;\ F^{-1}(u))\phi(u)du\ \text{and}\
S_2=\int_0^1\frac{(1-u)}{2nf(F^{-1}(u))}\int_{a_W}^{F^{-1}(u)}\frac{dW^*(x)}{C^2(x)}\phi(u)du.
\end{equation*}
We can write
\begin{align*}
&\sup_y|P(\sqrt{n}\sigma^{-1}(\widehat{M}_{Prod}-M_\phi)\leq y)-E_{n2}(y)| \\
&=\sup_y|P(\sqrt{n}\sigma^{-1}S_1\leq y-\sqrt{n}\sigma^{-1}S_2)-E_{n2}(y)| \\
&=\sup_y|P(\sqrt{n}\sigma^{-1}S_1\leq y)-E_{n1}(y)|+\sup_y|E_{n1}(y-\sqrt{n}\sigma^{-1}S_2)-E_{n2}(y)| \\
&=\sup_y|P(\sqrt{n}\sigma_{01}^{-1}S_1\leq y)-E_{n1}(y)|+\sup_y|E_{n1}(y-\sqrt{n}\sigma_{01}^{-1}S_2)-E_{n2}(y)| \\
&=A+B
\end{align*}

Now, using Lemma \ref{lem3.7} we can say that $A$ is $o(n^{-1/2})$. Clearly,
\begin{eqnarray*}
B&=&\sup_y|E_{n1}(y-\sqrt{n}\sigma_{01}^{-1}S_2)-E_{n2}(y)| \\
&=&\sup_y\Big|E_{n1}(y)-\Big[\frac{d}{dy}E_{n1}(y)\Big]\frac{1}{2\sqrt{n}}\sigma_{01}^{-1}\int_0^1\frac{(1-u)}{f(F^{-1}(u))}\int_{a_W}^{F^{-1}(u)}\frac{dW^*(x)}{C^2(x)}\phi(u)du-E_{n2}(y)\Big|+o(n^{-1/2}) \\
&=&o(n^{-1/2}),
\end{eqnarray*}
which completes the proof.

\noindent\textbf{Proof of Theorem \ref{theo3.9}:} Similar to Theorem (\ref{theo6.5}), we have
\begin{align}\label{Mp}\widehat{M}_{Prod}^*-\widehat{M}_{Prod}=S^*+\alpha_n^*, \end{align}
where $$S=-\int_0^1\frac{1}{f(F^{-1}(u))}\Big[\frac{(1-u)}{n^2}\sum_{i<j}h_1(V_i;V_j;F^{-1}(u))-\frac{1-u}{2n}\int_{a_W}^{F^{-1}(u)}\frac{dW^*(x)}{C^2(x)}\Big]\phi(u)du$$ and $S^*$ and $\alpha_n^*$ are the bootstrap analogs of $S$ and $\alpha_n$. A similar argument to the proof of Lemma 3 in the Appendix of Chen and Lo (1996) yields that

\begin{equation}\label{T*}\sup_y|P(\sqrt{n}T^*\leq y)-E_{n1}^*(y)|=o(n^{-1/2}),\ a.s.,\end{equation}
where $T=\sigma_{01}^{-1}\frac{1}{n^2}\sum_{i<j}h_1(V_i,V_j;F^{-1}(x))$ and $T^*$ and $E_{n1}^*(y)$ are the bootstrap analogs of $T$ and $E_{n1}(y)$. Similar to the proof of Theorem (\ref{theo6.4}) \cite{wang06}, we have with probability 1

\begin{align}\label{al}P(\sqrt{n}|\alpha_n^*|>cn^{-1/2}\log^{(-1)}{n})=o(n^{-1/2}). \end{align}
Hence, combining equation (\ref{Mp}), (\ref{T*}), and (\ref{al}) together with Lemma (\ref{lem3.6}) yield

\begin{equation*}
\sup_y|P^*(\sqrt{n}\sigma^{-1}(\widehat{M}_{Prod}^*-\widehat{M}_{Prod})\leq y)-E_{n2}^*(y)|=o(n^{-1/2}),\ a.s.
\end{equation*}
Now, it remains to prove that
\begin{equation}\label{En2}
\sup_y|E_{n2}(y)-E_{n2}^*(y)|=o(n^{-1/2}),\ a.s.
\end{equation}
Notice that
\begin{equation*}\sigma_{01}^{*^2}=\int_{a_W}^{F^{-1}(x)}\frac{dW^{*^*}(t)}{C^{*^2}(t)}=\int_{a_W}^{F^{-1}(x)}\frac{dW_n^{*}(t)}{C_n^{2}(t)},\\
\int_{a_W}^{F^{-1}(x)}\frac{dW^{*^*}(t)}{C^{*^3}(t)}=\int_{a_W}^{F^{-1}(x)}\frac{dW_n^{*}(t)}{C_n^{3}(t)}. \end{equation*}

Hence, $\kappa_3^*\xrightarrow{\text{p}}\kappa_3$. This proves (\ref{En2}). The proof of Theorem (\ref{theo3.9}) is thus complete.

\end{appendices}

\bibliography{sn-bibliography}


\begin{thebibliography}{38}
\ifx \bisbn   \undefined \def \bisbn  #1{ISBN #1}\fi
\ifx \binits  \undefined \def \binits#1{#1}\fi
\ifx \bauthor  \undefined \def \bauthor#1{#1}\fi
\ifx \batitle  \undefined \def \batitle#1{#1}\fi
\ifx \bjtitle  \undefined \def \bjtitle#1{#1}\fi
\ifx \bvolume  \undefined \def \bvolume#1{\textbf{#1}}\fi
\ifx \byear  \undefined \def \byear#1{#1}\fi
\ifx \bissue  \undefined \def \bissue#1{#1}\fi
\ifx \bfpage  \undefined \def \bfpage#1{#1}\fi
\ifx \blpage  \undefined \def \blpage #1{#1}\fi
\ifx \burl  \undefined \def \burl#1{\textsf{#1}}\fi
\ifx \doiurl  \undefined \def \doiurl#1{\url{https://doi.org/#1}}\fi
\ifx \betal  \undefined \def \betal{\textit{et al.}}\fi
\ifx \binstitute  \undefined \def \binstitute#1{#1}\fi
\ifx \binstitutionaled  \undefined \def \binstitutionaled#1{#1}\fi
\ifx \bctitle  \undefined \def \bctitle#1{#1}\fi
\ifx \beditor  \undefined \def \beditor#1{#1}\fi
\ifx \bpublisher  \undefined \def \bpublisher#1{#1}\fi
\ifx \bbtitle  \undefined \def \bbtitle#1{#1}\fi
\ifx \bedition  \undefined \def \bedition#1{#1}\fi
\ifx \bseriesno  \undefined \def \bseriesno#1{#1}\fi
\ifx \blocation  \undefined \def \blocation#1{#1}\fi
\ifx \bsertitle  \undefined \def \bsertitle#1{#1}\fi
\ifx \bsnm \undefined \def \bsnm#1{#1}\fi
\ifx \bsuffix \undefined \def \bsuffix#1{#1}\fi
\ifx \bparticle \undefined \def \bparticle#1{#1}\fi
\ifx \barticle \undefined \def \barticle#1{#1}\fi
\bibcommenthead
\ifx \bconfdate \undefined \def \bconfdate #1{#1}\fi
\ifx \botherref \undefined \def \botherref #1{#1}\fi
\ifx \url \undefined \def \url#1{\textsf{#1}}\fi
\ifx \bchapter \undefined \def \bchapter#1{#1}\fi
\ifx \bbook \undefined \def \bbook#1{#1}\fi
\ifx \bcomment \undefined \def \bcomment#1{#1}\fi
\ifx \oauthor \undefined \def \oauthor#1{#1}\fi
\ifx \citeauthoryear \undefined \def \citeauthoryear#1{#1}\fi
\ifx \endbibitem  \undefined \def \endbibitem {}\fi
\ifx \bconflocation  \undefined \def \bconflocation#1{#1}\fi
\ifx \arxivurl  \undefined \def \arxivurl#1{\textsf{#1}}\fi
\csname PreBibitemsHook\endcsname

\bibitem[\protect\citeauthoryear{Acerbi}{2002}]{acerbi02}
\begin{barticle}
\bauthor{\bsnm{Acerbi}, \binits{C.}}:
\batitle{Spectral measures of risk: A coherent representation of subjective
  risk aversion}.
\bjtitle{Journal of Banking \& Finance}
\bvolume{26}(\bissue{7}),
\bfpage{1505}--\blpage{1518}
(\byear{2002})
\end{barticle}
\endbibitem

\bibitem[\protect\citeauthoryear{Artzner et~al.}{1999}]{Artz99}
\begin{barticle}
\bauthor{\bsnm{Artzner}, \binits{P.}},
\bauthor{\bsnm{Delbaen}, \binits{F.}},
\bauthor{\bsnm{Eber}, \binits{J.-M.}},
\bauthor{\bsnm{Heath}, \binits{D.}}:
\batitle{Coherent measures of risk}.
\bjtitle{Mathematical finance}
\bvolume{9}(\bissue{3}),
\bfpage{203}--\blpage{228}
(\byear{1999})
\end{barticle}
\endbibitem

\bibitem[\protect\citeauthoryear{Brazauskas and
  Kleefeld}{2016}]{brazauskas2016}
\begin{barticle}
\bauthor{\bsnm{Brazauskas}, \binits{V.}},
\bauthor{\bsnm{Kleefeld}, \binits{A.}}:
\batitle{Modeling severity and measuring tail risk of norwegian fire claims}.
\bjtitle{North American Actuarial Journal}
\bvolume{20}(\bissue{1}),
\bfpage{1}--\blpage{16}
(\byear{2016})
\end{barticle}
\endbibitem

\bibitem[\protect\citeauthoryear{Cotter and Dowd}{2006}]{cotter06}
\begin{barticle}
\bauthor{\bsnm{Cotter}, \binits{J.}},
\bauthor{\bsnm{Dowd}, \binits{K.}}:
\batitle{Extreme spectral risk measures: an application to futures
  clearinghouse margin requirements}.
\bjtitle{Journal of Banking \& Finance}
\bvolume{30}(\bissue{12}),
\bfpage{3469}--\blpage{3485}
(\byear{2006})
\end{barticle}
\endbibitem

\bibitem[\protect\citeauthoryear{Chang}{1991}]{chang91}
\begin{barticle}
\bauthor{\bsnm{Chang}, \binits{M.N.}}:
\batitle{Edgeworth expansion for the kaplan-meier estimato}.
\bjtitle{Communications in statistics-theory and methods}
\bvolume{20}(\bissue{8}),
\bfpage{2479}--\blpage{2494}
(\byear{1991})
\end{barticle}
\endbibitem

\bibitem[\protect\citeauthoryear{Cheng et~al.}{2016}]{cheng2016}
\begin{barticle}
\bauthor{\bsnm{Cheng}, \binits{J.-Y.}},
\bauthor{\bsnm{Huang}, \binits{S.-C.}},
\bauthor{\bsnm{Tzeng}, \binits{S.-J.}}:
\batitle{Quantile regression methods for left-truncated and right-censored
  data}.
\bjtitle{Journal of Statistical Computation and Simulation}
\bvolume{86}(\bissue{3}),
\bfpage{443}--\blpage{459}
(\byear{2016})
\end{barticle}
\endbibitem

\bibitem[\protect\citeauthoryear{Calder{\'\i}n-Ojeda et~al.}{2023}]{Ojeda2023}
\begin{barticle}
\bauthor{\bsnm{Calder{\'\i}n-Ojeda}, \binits{E.}},
\bauthor{\bsnm{G{\'o}mez-D{\'e}niz}, \binits{E.}},
\bauthor{\bsnm{V{\'a}zquez-Polo}, \binits{F.J.}}:
\batitle{Conditional tail expectation and premium calculation under asymmetric
  loss}.
\bjtitle{Axioms}
\bvolume{12}(\bissue{5}),
\bfpage{496}
(\byear{2023})
\end{barticle}
\endbibitem

\bibitem[\protect\citeauthoryear{Cheng~Wang}{1987}]{wang1987}
\begin{barticle}
\bauthor{\bsnm{Cheng~Wang}, \binits{M.}}:
\batitle{Product limit estimates: a generalized maximum likelihood study}.
\bjtitle{Communications in Statistics-Theory and Methods}
\bvolume{16}(\bissue{11}),
\bfpage{3117}--\blpage{3132}
(\byear{1987})
\end{barticle}
\endbibitem

\bibitem[\protect\citeauthoryear{Dowd and Blake}{2006}]{dowd06}
\begin{barticle}
\bauthor{\bsnm{Dowd}, \binits{K.}},
\bauthor{\bsnm{Blake}, \binits{D.}}:
\batitle{After var: the theory, estimation, and insurance applications of
  quantile-based risk measures}.
\bjtitle{Journal of Risk and Insurance}
\bvolume{73}(\bissue{2}),
\bfpage{193}--\blpage{229}
(\byear{2006})
\end{barticle}
\endbibitem

\bibitem[\protect\citeauthoryear{Delbaen}{2002}]{del02}
\begin{bchapter}
\bauthor{\bsnm{Delbaen}, \binits{F.}}:
\bctitle{Coherent risk measures on general probability spaces}.
In: \bbtitle{Advances in Finance and Stochastics},
pp. \bfpage{1}--\blpage{37}.
\bpublisher{Springer}, \blocation{???}
(\byear{2002})
\end{bchapter}
\endbibitem

\bibitem[\protect\citeauthoryear{Efron}{1979}]{efron79}
\begin{barticle}
\bauthor{\bsnm{Efron}, \binits{B.}}:
\batitle{Bootstrap method: another look at the jackknife.}
\bjtitle{The Annals of Statistics}
\bvolume{7}(\bissue{1}),
\bfpage{1}--\blpage{26}
(\byear{1979})
\end{barticle}
\endbibitem

\bibitem[\protect\citeauthoryear{Ergashev et~al.}{2016}]{Erg16}
\begin{barticle}
\bauthor{\bsnm{Ergashev}, \binits{B.}},
\bauthor{\bsnm{Pavlikov}, \binits{K.}},
\bauthor{\bsnm{Uryasev}, \binits{S.}},
\bauthor{\bsnm{Sekeris}, \binits{E.}}:
\batitle{Estimation of truncated data samples in operational risk modeling}.
\bjtitle{Journal of Risk and Insurance}
\bvolume{83}(\bissue{3}),
\bfpage{613}--\blpage{640}
(\byear{2016})
\end{barticle}
\endbibitem

\bibitem[\protect\citeauthoryear{G{\"u}rler et~al.}{1993}]{Gurler1993}
\begin{barticle}
\bauthor{\bsnm{G{\"u}rler}, \binits{{\"U}.}},
\bauthor{\bsnm{Stute}, \binits{W.}},
\bauthor{\bsnm{Wang}, \binits{J.-L.}}:
\batitle{Weak and strong quantile representations for randomly truncated data
  with applications}.
\bjtitle{Statistics \& probability letters}
\bvolume{17}(\bissue{2}),
\bfpage{139}--\blpage{148}
(\byear{1993})
\end{barticle}
\endbibitem

\bibitem[\protect\citeauthoryear{Gu}{1995}]{gu1995}
\begin{barticle}
\bauthor{\bsnm{Gu}, \binits{M.}}:
\batitle{Convergence of increments for cumulative hazard function in a mixed
  censorship-truncation model with application to hazard estimators}.
\bjtitle{Statistics \& probability letters}
\bvolume{23}(\bissue{2}),
\bfpage{135}--\blpage{139}
(\byear{1995})
\end{barticle}
\endbibitem

\bibitem[\protect\citeauthoryear{Gijbels and Wang}{1993}]{gijbels1993}
\begin{barticle}
\bauthor{\bsnm{Gijbels}, \binits{I.}},
\bauthor{\bsnm{Wang}, \binits{J.-L.}}:
\batitle{Strong representations of the survival function estimator for
  truncated and censored data with applications}.
\bjtitle{Journal of Multivariate analysis}
\bvolume{47}(\bissue{2}),
\bfpage{210}--\blpage{229}
(\byear{1993})
\end{barticle}
\endbibitem

\bibitem[\protect\citeauthoryear{Heras et~al.}{2012}]{heras12}
\begin{barticle}
\bauthor{\bsnm{Heras}, \binits{A.}},
\bauthor{\bsnm{Balbas}, \binits{B.}},
\bauthor{\bsnm{Vilar}, \binits{J.L.}}:
\batitle{Conditional tail expectation and premium calculation}.
\bjtitle{ASTIN Bulletin: The Journal of the IAA}
\bvolume{42}(\bissue{1}),
\bfpage{325}--\blpage{342}
(\byear{2012})
\end{barticle}
\endbibitem

\bibitem[\protect\citeauthoryear{Henryk and Silvia}{2008}]{gzyland08}
\begin{barticle}
\bauthor{\bsnm{Henryk}, \binits{G.}},
\bauthor{\bsnm{Silvia}, \binits{M.}}:
\batitle{On a relationship between distorted and spectral risk measures}.
\bjtitle{Revista de econom$\acute{i}$a financiera}
\bvolume{15},
\bfpage{8}--\blpage{21}
(\byear{2008})
\end{barticle}
\endbibitem

\bibitem[\protect\citeauthoryear{Kaiser and Brazauskas}{2006}]{kaiser2006}
\begin{barticle}
\bauthor{\bsnm{Kaiser}, \binits{T.}},
\bauthor{\bsnm{Brazauskas}, \binits{V.}}:
\batitle{Interval estimation of actuarial risk measures}.
\bjtitle{North American Actuarial Journal}
\bvolume{10}(\bissue{4}),
\bfpage{249}--\blpage{268}
(\byear{2006})
\end{barticle}
\endbibitem

\bibitem[\protect\citeauthoryear{Klugman et~al.}{2019}]{klugman19}
\begin{bbook}
\bauthor{\bsnm{Klugman}, \binits{S.A.}},
\bauthor{\bsnm{Panjer}, \binits{H.H.}},
\bauthor{\bsnm{Willmot}, \binits{G.E.}}:
\bbtitle{Loss Models: from Data to Decisions},
\bedition{Fifth edition} edn.
\bpublisher{John Wiley \& Sons},
\blocation{New Jersey}
(\byear{2019})
\end{bbook}
\endbibitem

\bibitem[\protect\citeauthoryear{Liang et~al.}{2015}]{liang2015}
\begin{barticle}
\bauthor{\bsnm{Liang}, \binits{H.}},
\bauthor{\bsnm{Li}, \binits{D.}},
\bauthor{\bsnm{Miao}, \binits{T.}}:
\batitle{Conditional quantile estimation with truncated, censored and dependent
  data}.
\bjtitle{Chinese Annals of Mathematics, Series B}
\bvolume{36}(\bissue{6}),
\bfpage{969}--\blpage{990}
(\byear{2015})
\end{barticle}
\endbibitem

\bibitem[\protect\citeauthoryear{Lai and Ying}{1991}]{lai1991}
\begin{botherref}
\oauthor{\bsnm{Lai}, \binits{T.L.}},
\oauthor{\bsnm{Ying}, \binits{Z.}}:
Estimating a distribution function with truncated and censored data.
The Annals of Statistics,
417--442
(1991)
\end{botherref}
\endbibitem

\bibitem[\protect\citeauthoryear{Nadarajah and Bakar}{2015}]{nadarajah2015}
\begin{barticle}
\bauthor{\bsnm{Nadarajah}, \binits{S.}},
\bauthor{\bsnm{Bakar}, \binits{S.A.}}:
\batitle{New folded models for the log-transformed norwegian fire claim data}.
\bjtitle{Communications in Statistics-Theory and Methods}
\bvolume{44}(\bissue{20}),
\bfpage{4408}--\blpage{4440}
(\byear{2015})
\end{barticle}
\endbibitem

\bibitem[\protect\citeauthoryear{Pichler}{2015}]{pichler2015}
\begin{barticle}
\bauthor{\bsnm{Pichler}, \binits{A.}}:
\batitle{Premiums and reserves, adjusted by distortions}.
\bjtitle{Scandinavian Actuarial Journal}
\bvolume{2015}(\bissue{4}),
\bfpage{332}--\blpage{351}
(\byear{2015})
\end{barticle}
\endbibitem

\bibitem[\protect\citeauthoryear{Shi et~al.}{2018}]{Shi2018}
\begin{barticle}
\bauthor{\bsnm{Shi}, \binits{J.}},
\bauthor{\bsnm{Ma}, \binits{H.}},
\bauthor{\bsnm{Zhou}, \binits{Y.}}:
\batitle{The nonparametric quantile estimation for length-biased and
  right-censored data}.
\bjtitle{Statistics \& Probability Letters}
\bvolume{134},
\bfpage{150}--\blpage{158}
(\byear{2018})
\end{barticle}
\endbibitem

\bibitem[\protect\citeauthoryear{Stute}{1993}]{stute1993}
\begin{botherref}
\oauthor{\bsnm{Stute}, \binits{W.}}:
Almost sure representations of the product-limit estimator for truncated data.
The Annals of Statistics,
146--156
(1993)
\end{botherref}
\endbibitem

\bibitem[\protect\citeauthoryear{Sun}{1997}]{sun1997}
\begin{barticle}
\bauthor{\bsnm{Sun}, \binits{L.}}:
\batitle{Bandwidth choice for hazard rate estimators from left truncated and
  right censored data}.
\bjtitle{Statistics \& probability letters}
\bvolume{36}(\bissue{2}),
\bfpage{101}--\blpage{114}
(\byear{1997})
\end{barticle}
\endbibitem

\bibitem[\protect\citeauthoryear{Sun and Zhou}{1998}]{sun1998}
\begin{barticle}
\bauthor{\bsnm{Sun}, \binits{L.}},
\bauthor{\bsnm{Zhou}, \binits{Y.}}:
\batitle{Sequential confidence bands for densities under truncated and censored
  data}.
\bjtitle{Statistics \& probability letters}
\bvolume{40}(\bissue{1}),
\bfpage{31}--\blpage{41}
(\byear{1998})
\end{barticle}
\endbibitem

\bibitem[\protect\citeauthoryear{Tsai et~al.}{1987}]{tsai1987}
\begin{barticle}
\bauthor{\bsnm{Tsai}, \binits{W.-Y.}},
\bauthor{\bsnm{Jewell}, \binits{N.P.}},
\bauthor{\bsnm{Wang}, \binits{M.-C.}}:
\batitle{A note on the product-limit estimator under right censoring and left
  truncation}.
\bjtitle{Biometrika}
\bvolume{74}(\bissue{4}),
\bfpage{883}--\blpage{886}
(\byear{1987})
\end{barticle}
\endbibitem

\bibitem[\protect\citeauthoryear{Tsukahara}{2009}]{Tsu09b}
\begin{barticle}
\bauthor{\bsnm{Tsukahara}, \binits{H.}}:
\batitle{One-parameter families of distortion risk measures}.
\bjtitle{Mathematical Finance: An International Journal of Mathematics,
  Statistics and Financial Economics}
\bvolume{19}(\bissue{4}),
\bfpage{691}--\blpage{705}
(\byear{2009})
\end{barticle}
\endbibitem

\bibitem[\protect\citeauthoryear{Upretee}{2020}]{upretee2020}
\begin{botherref}
\oauthor{\bsnm{Upretee}, \binits{S.}}:
Estimating distortion risk measures under truncated and censored data
  scenarios.
PhD thesis,
The University of Wisconsin-Milwaukee
(2020)
\end{botherref}
\endbibitem

\bibitem[\protect\citeauthoryear{UZUNOG{\={}}~ULLARI and
  Wang}{1992}]{uzunog1992}
\begin{barticle}
\bauthor{\bsnm{UZUNOG{\={}}~ULLARI}, \binits{{\"U}.}},
\bauthor{\bsnm{Wang}, \binits{J.-L.}}:
\batitle{A comparison of hazard rate estimators for left truncated and right
  censored data}.
\bjtitle{Biometrika}
\bvolume{79}(\bissue{2}),
\bfpage{297}--\blpage{310}
(\byear{1992})
\end{barticle}
\endbibitem

\bibitem[\protect\citeauthoryear{Wang}{1996}]{wang1996}
\begin{barticle}
\bauthor{\bsnm{Wang}, \binits{S.}}:
\batitle{Premium calculation by transforming the layer premium density}.
\bjtitle{ASTIN Bulletin: The Journal of the IAA}
\bvolume{26}(\bissue{1}),
\bfpage{71}--\blpage{92}
(\byear{1996})
\end{barticle}
\endbibitem

\bibitem[\protect\citeauthoryear{Wang and Jing}{2006}]{wang06}
\begin{barticle}
\bauthor{\bsnm{Wang}, \binits{Q.}},
\bauthor{\bsnm{Jing}, \binits{B.-Y.}}:
\batitle{Edgeworth expansion and bootstrap approximation for studentized
  product-limit estimator with truncated and censored data}.
\bjtitle{Communications in Statistics-Theory and Methods}
\bvolume{35}(\bissue{4}),
\bfpage{609}--\blpage{623}
(\byear{2006})
\end{barticle}
\endbibitem

\bibitem[\protect\citeauthoryear{Woodroofe}{1985}]{wood1985}
\begin{barticle}
\bauthor{\bsnm{Woodroofe}, \binits{M.}}:
\batitle{Estimating a distribution function with truncated data}.
\bjtitle{The Annals of Statistics}
\bvolume{13}(\bissue{1}),
\bfpage{163}--\blpage{177}
(\byear{1985})
\end{barticle}
\endbibitem

\bibitem[\protect\citeauthoryear{Wang and Young}{1998}]{wang98}
\begin{barticle}
\bauthor{\bsnm{Wang}, \binits{S.S.}},
\bauthor{\bsnm{Young}, \binits{V.R.}}:
\batitle{Risk-adjusted credibility premiums using distorted probabilities}.
\bjtitle{Scandinavian Actuarial Journal}
\bvolume{1998}(\bissue{2}),
\bfpage{143}--\blpage{165}
(\byear{1998})
\end{barticle}
\endbibitem

\bibitem[\protect\citeauthoryear{Zhou}{1996}]{zhou1996}
\begin{barticle}
\bauthor{\bsnm{Zhou}, \binits{Y.}}:
\batitle{A note on the tjw product-limit estimator for truncated and censored
  data}.
\bjtitle{Statistics \& Probability Letters}
\bvolume{26}(\bissue{4}),
\bfpage{381}--\blpage{387}
(\byear{1996})
\end{barticle}
\endbibitem

\bibitem[\protect\citeauthoryear{Zhou et~al.}{2000}]{zhou2000}
\begin{botherref}
\oauthor{\bsnm{Zhou}, \binits{X.}},
\oauthor{\bsnm{Sun}, \binits{L.}},
\oauthor{\bsnm{Ren}, \binits{H.}}:
Quantile estimation for left truncated and right censored data.
Statistica Sinica,
1217--1229
(2000)
\end{botherref}
\endbibitem

\bibitem[\protect\citeauthoryear{Zhou and Yip}{1999}]{zhou1999}
\begin{barticle}
\bauthor{\bsnm{Zhou}, \binits{Y.}},
\bauthor{\bsnm{Yip}, \binits{P.S.}}:
\batitle{A strong representation of the product-limit estimator for left
  truncated and right censored data}.
\bjtitle{Journal of Multivariate Analysis}
\bvolume{69}(\bissue{2}),
\bfpage{261}--\blpage{280}
(\byear{1999})
\end{barticle}
\endbibitem

\end{thebibliography}

\end{document}